\documentclass[journal]{new-aiaa}
\usepackage[utf8]{inputenc}

\usepackage{graphicx}
\usepackage{amsmath}
\usepackage[version=4]{mhchem}
\usepackage{longtable,tabularx}
\newcommand{\pderiv}[2]{\frac{\partial #1}{\partial #2}}

\title{Analytical framework for space debris collision avoidance maneuver design
\footnote[0]{A first version of this work was presented as Paper AAS 18-357 at the 2018 AAS/AIAA Astrodynamics Specialist Conference, 19-23 August 2018, Snowbird, UT.}}

\author{Juan Luis Gonzalo \footnote{Postdoctoral research fellow, Department of Aerospace Science and Technology, Via la Masa 34, 20156 Milan, Italy. AIAA Member}, Camilla Colombo \footnote{Associate professor, Department of Aerospace Science and Technology, Via la Masa 34, 20156 Milan, Italy} and Pierluigi Di Lizia \footnote{Assistant professor, Department of Aerospace Science and Technology, Via la Masa 34, 20156 Milan, Italy. AIAA Member}}
\affil{Polytechnic University of Milan, 20156 Milan, Italy}

\begin{document}

\maketitle

\begin{abstract}
An analytical formulation for collision avoidance maneuvers involving a spacecraft and a space debris is presented, including solutions for the maximum deviation and minimum collision probability cases. Gauss' planetary equations and relative motion equations are used to map maneuvers at a given time to displacements at the predicted close approach. The model is then extended to map changes in state between two times, allowing one to propagate covariance matrices. The analytical formulation reduces the optimization problem to an eigenproblem, both for maximum deviation and minimum collision probability. Two maximum deviation cases, total deviation and impact parameter, are compared for a large set of spacecraft-debris conjunction geometries derived from European Space Agency’s Meteoroid and Space Debris Terrestrial Environment Reference (MASTER-2009) model. Moreover, the maximum impact parameter and minimum collision probability maneuvers are compared assuming covariances known at the maneuver time, to evaluate the net effect of lead time in collision probability. In all cases, solutions are analyzed in the b-plane to leverage its natural separation of phasing and geometry change effects. Both uncertainties and maximum deviation grow along the time axis for long lead times, limiting the reduction in collision probability.
\end{abstract}

\section*{Nomenclature}

{\renewcommand\arraystretch{1.0}
\noindent\begin{longtable*}{@{}l @{\quad=\quad} l@{}}
$\boldsymbol{\alpha}$ & Keplerian elements of the orbit $[ \, a \; e \; i \; \omega \; \Omega \; M \, ]$, $\mathrm{km}$ and $\mathrm{deg}$ or $\mathrm{rad}$\\
$\Delta t$ & Lead time for the collision avoidance maneuver, $\mathrm{s}$\\
$\Delta V$ & Relative velocity between debris and spacecraft at the close approach, $\mathrm{km}/\mathrm{s}$\\
$\delta \mathbf{b}^*$ & Deviation in the b-plane after maneuver, $\mathrm{km}$\\
$\delta \mathbf{r}$ & Total deviation after maneuver, $\mathrm{km}$\\
$\delta \mathbf{v}$ & Impulsive maneuver, $\mathrm{km}/\mathrm{s}$\\
$\delta x$ & Change in generic magnitude $x$\\
$\zeta$ & Coordinate in the time axis of the b-plane, $\mathrm{km}$\\
$\mu$ & Gravitational parameter of the primary, $\mathrm{km}^3/\mathrm{s}^2$\\
$\xi$ & Coordinate in the geometry axis of the b-plane, $\mathrm{km}$\\
$\rho_{xy}$ & Correlation for generic variables $x$ and $y$\\
$\sigma_x$ & Covariance for generic variable $x$\\
$\Omega$ & Right ascension of the ascending node, $\mathrm{deg}$ or $\mathrm{rad}$\\
$\omega$ & Argument of perigee, $\mathrm{deg}$ or $\mathrm{rad}$\\
$\mathbf{A}_r$ & Matrix form of the linearized relative motion equations (only for position)\\
$\mathbf{A}$ & Matrix form of the linearized relative motion equations\\
$a$ & Semi-major axis, $\mathrm{km}$\\
$b^*$ & Impact parameter, $\mathrm{km}$\\
$\mathbf{C}$ & Covariance matrix\\
$e$ & Eccentricity\\
$f$ & True anomaly, $\mathrm{deg}$ or $\mathrm{rad}$\\
$\mathbf{G}$ & Jacobian of Keplerian elements with respect to Cartesian state vector\\
$\mathbf{G}_v$ & Matrix form of the Gauss' planetary equations\\
$\mathbf{h}$ & Angular momentum, $\mathrm{km}^2/\mathrm{s}$\\
$i$ & Inclination, $\mathrm{deg}$ or $\mathrm{rad}$\\
$M$ & Mean anomaly, $\mathrm{deg}$ or $\mathrm{rad}$\\
$P$ & Collision probability\\
$p$ & Parameter of the orbit, $\mathrm{km}$\\
$\mathbf{r}$ & Position vector, $\mathrm{km}$\\
$r_A$ & Radius of the spacecraft-debris combined envelope, $\mathrm{km}$\\
$\mathbf{s}$ & Cartesian state vector, $\mathrm{km}$ and $\mathrm{km}/\mathrm{s}$\\
$\mathbf{T}$ & State transition matrix relating $\delta\mathbf{v}$ at $t_\text{CAM}$ with $\delta\mathbf{r}$ at $t_\text{CA}$\\
$T$ & Orbital period of the spacecraft, $\mathrm{s}$\\
$t$ & Time, $\mathrm{s}$\\
$t_\mathrm{CA}$ & Time at the close approach, $\mathrm{s}$\\
$t_\mathrm{CAM}$ & Time at which the collision avoidance maneuver is performed, $\mathrm{s}$\\
$\mathbf{v}$ & Velocity, $\mathrm{km}/\mathrm{s}$\\
$\mathbf{Z}$ & State transition matrix relating $\delta\mathbf{v}$ at $t_\text{CAM}$ with deviation in the b-plane at $t_\text{CA}$

\end{longtable*}}

\section{Introduction}

\lettrine{T}{he} need for and complexity of collision avoidance activities between active spacecraft and debris (or other spacecraft) has experienced a notable increase in the last couple decades, due to the growing number of satellites in orbit and significant fragmentations events (most notably, Fengyun-1C in 2007~\cite{johnson2008characteristics}, and Iridium-33/Cosmos-2251 in 2009 \cite{kelso2009analysis}). The proliferation of objects in Earth orbit already poses a critical threat to the safe and sustainable use of space, and it is expected to keep increasing furthered by recent developments in the space sector, such as new launch companies driving down the access cost to space, the popularization of small, cube and nanosats as affordable yet flexible platforms, and the large constellations being proposed both by incumbent companies and startups. Several international efforts are being undertaken to tackle this issue, such as the Inter-Agency Space Debris Coordination Committee (IADC) and its space debris mitigation guidelines: a maximum lifetime of 25 years for objects in low Earth orbit, and the relocation to graveyard orbits for objects in geostationary Earth orbit. However, complying with these guidelines introduces additional costs and complexity. For instance, if a mitigation maneuver is to be performed using the satellite thrusters the additional operational time and propellant requirements have to be considered. Furthermore, this may not be a feasible option for all kinds of platforms. A cost-effective alternative for reducing the de-orbiting time of small satellites can be the use of passive end-of-life de-orbiting methods, such as drag or solar sails and electrodynamic tethers. On the downside, their relatively large cross-sectional area appreciably increases the risk of collision with other spacecraft or space debris during the deorbiting phase~\cite{colombo2017drag,colombo2018effects,colombo2019long}.

Space situational awareness and collision avoidance activities for active satellites are also hindered by the fact that debris and spacecraft tracking, collision risk assessment, and spacecraft operations are in many cases performed by different entities. The U.S. Strategic Command (USSTRATCOM) provides a publicly available catalog of space objects around Earth. The catalog is constructed using observations from the Joint Space Operations Center (JSoPC) space surveillance network, and orbit determination and propagation algorithms based on the simplified general perturbations (SGP) model~\cite{hoots2004history,vallado2006revisiting}. Public data from the catalog is made available in the form of two-line elements (TLEs). TLEs have a limited accuracy, and come without information about the uncertainties in the orbit determination. This implies important limitations to the use of catalog information for Close Approach (CA) prediction, requiring to improve the accuracy or estimate the uncertainties with additional observations or numerical techniques \cite{flohrer2008assesment,frueh2012accuracy,morselli2014high}.
The accuracy of CA-related information improved substantially with the introduction of Conjunction Data Messages (CDMs) and Conjunction Summary Messages (CSMs)~\cite{bird2010sharing}, which provide detailed orbit information along with the full covariance quantifying the uncertainties in orbit determination. However, this information is only provided to the spacecraft operator, and only when a possible CA is detected. The number of CDM/CSM generated by USSTRATCOM and posted to Space-Track.org per day can be in the thousands.

The publicly available information on how different entities carry out their CA monitoring and Collision Avoidance Maneuver (CAM) planning activities for the spacecraft they operate is limited. A detailed overview of the methodology and tools used by European Space Agency’s (ESA) Space Debris Office (SDO), which provides operational collision avoidance services for ESA missions and third parties, is offered by Braun et al. in~\cite{braun2016operational}. Two key aspects are highlighted in this reference. On the one hand, the evolution of the information available to operators. On the other hand, the introduction of new computational tools to efficiently handling the increasing amount of information. To address these issues, the SDO has integrated a variety of tools~\cite{braun2016operational,flohrer2011operational}. It maintains the Database Information System Characterizing Objects in Space (DISCOS) database~\cite{flohrer2013discos,klinkrad1991discos}, providing physical information about objects on orbit. Debris analysis is performed using the Debris Risk Assessment and Mitigation Analysis (DRAMA) software suite~\cite{martin2005introducing}, including the Meteoroid and Space Debris Terrestrial Environment Reference (MASTER) \cite{flegel2009master} and Assessment of Risk Event Statistics (ARES) tools. The former provides detailed information on the space debris population, while the latter computes statistical information such as estimates on annual CAMs required for a given orbital region and desired level of risk. Collision risk is evaluated with Collision Risk Assessment Software (CRASS) \cite{alarconrodriguez2004development,klinkrad2005collision}, merging information from different sources (TLEs, CDMs/CSMs, ephemeris from operators) and maintaining its own database~\cite{braun2016operational}. SDO’s capabilities were recently extended with the introduction of Collision Risk Assessment and Avoidance Manoeuvre (CORAM) \cite{sanchez2013collision,pulido2014coram,pulido2014esa}, which performs extended risk assessment and collision avoidance maneuver planning and optimization through its Collision Risk Computation Software (CORCOS) and Collision Avoidance Manoeuvre Optimization Software (CAMOS) tools, respectively.

Another important aspect is the choice of a collision probability threshold to operate. Setting a threshold too low would increase both the required effort from the operations team and fuel consumption, whereas setting it too high can raise the risk to unacceptable levels. The accuracy improvements on the available information, as well as the introduction of more advanced software tools, has allowed to reduce these thresholds in recent times. Reference~\cite{braun2016operational} provides information about the evolution of the different events' trigger thresholds and post-maneuver target at the SDO depending on the screening method. Although the specific values would be mission-dependent, reference values are provided for the threshold to initiate the decision process ($P>{10}^{-4}$) and for the post-maneuver target ($P<{10}^{-6}$) as of 2013, using the CDM/CSM with full covariance information for screening.

Despite all the advances, many challenges remain for the future. The increasing space traffic will raise the need of CAMs by active satellites and the amount of data to be managed, an issue which could be tackled by the creation of Space Traffic Management protocols and entities~\cite{muelhaupt2019space}, analogously to current air traffic management activities. The United States has recently proposed to transfer all its civil Space Traffic Management-related activities to the Department of Commerce\footnote{Space Policy Directive-3, \url{https://www.whitehouse.gov/presidential-actions/space-policy-directive-3-national-space-traffic-management-policy/} [last accessed 20/11/2018]}, including many activities currently carried out by JSoPC such as maintaining the publicly releasable portion of the space object catalogue, and on-orbit  collision avoidance support services. Regarding space debris identification and tracking, the Space Fence system currently under development by Lockheed Martin\footnote{Source: \url{https://www.lockheedmartin.com/en-us/products/space-fence.html} [last accessed 08/02/2019]}  will notably improve the capabilities of the U.S. Air Force Space Surveillance Network, allowing to track debris smaller than the current 10 cm limit (down to 2-5 cm).

In recent years, a significant amount of research has been carried out on CAM modelling and optimization to advance the state of the art and address some of these challenges. Bombardelli proposed in~\cite{bombardelli2014analytical} an analytical formulation for impulsive CAMs based on the Dromo set of regularized orbital elements. Accurate and relatively simple expressions are proposed for the characterization of the relative dynamics in the b-plane, and a procedure for the design of maximum miss distance CAMs is derived. These results were later extended by Bombardelli and Hernando-Ayuso~\cite{bombardelli2015optimal} to include also the design of minimum collision probability CAMs. Vasile et al.~\cite{vasile2017artificial} proposed an artificial intelligence-based approach to assist in the decision-making process for space traffic management activities. By using a database of CAs and CAMs to train a machine learning algorithm, their approach accounts for the consequences of a maneuver (e.g. future CAs) when designing the CAM. Kim et al.~\cite{kim2012study} used genetic algorithms to tackle the case where several debris approach the satellite in a short period of time, identifying limitations in the use of tangential maneuvers to deal with this scenario. Furthermore, several authors are investigating the possibility to perform so-called just-in-time CAMs between two debris, using external actions such as lasers or clouds of gas \cite{bonnal2020just,manson2011orbital}.

This paper focuses on the analysis and design of optimal impulsive CAMs involving a spacecraft and a space debris using analytical methods. Naturally, the results also apply to CAMs between two active spacecraft provided that only one of them performs a maneuver. Analytical methods can prove very useful for applications requiring high computational efficiency, e.g. analyses over large sets of data or on-board applications. Following the procedure proposed by Vasile et al.~\cite{vasile2008optimal} for the optimal deflection of asteroids, the instantaneous change in orbital elements due to an impulsive maneuver is computed through Gauss' planetary equations. Because the change in orbital elements is typically small, the deviation at the CA can then be computed through linearized relative motion equations, leading to a linear model with a matrix depending on the nominal orbital elements of the deflected body and the lead time of the maneuver. Applying a previous result by Conway~\cite{conway2001near}, the maximum deviation optimal control problem can be reduced to an eigenproblem.

The present work has two main differences compared to the asteroid deflection model by Vasile et al.~\cite{vasile2008optimal} or its application to artificial intelligence-assisted CAM design in~\cite{vasile2017artificial}. On the one hand, the formulation is extended to the optimization of minimum collision probability CAMs following the method proposed by Bombardelli and Hernando-Ayuso~\cite{bombardelli2015optimal}. By using Chan's method for the computation of collision probabilities between two objects~\cite{chan2008spacecraft}, they reduce the optimal collision probability problem to other with the same structure as the maximum deviation one considered by Conway~\cite{conway2001near}, thus solvable as an eigenproblem. On the other hand, the formulation is extended to map changes in the state at a given time to changes in the state at the predicted CA. This is particularly useful for the analytical propagation of covariance matrices, with the limitation that no perturbations are included. Another minor difference is that the matrix form of the equations is rearranged, to separate the effects due to the maneuver, the coasting arc up to the CA, and the evaluation of the relative motion equations. This change eases the extension of the formulation to other force models currently under investigation~\cite{gonzalo2019semianalytical}.

The b-plane (see~\cite{opik1976interplanetary,kizner1961method}) will be used extensively for the analysis of the results, leveraging its separation of the displacements at the CA along a time axis, associated with phasing maneuvers, and a geometry axis, associated to changes in the shape of the orbit. Furthermore, two different maximum deviation CAMs are considered: maximum total deviation and maximum impact parameter in the b-plane. Both CAMs are compared through an extensive sensitivity analysis over the possible conjunction geometries, derived using statistical data from ESA's MASTER-2009 for the space debris population. Special attention is paid to assessing the qualitative and quantitative effects of orbit eccentricity, conjunction geometry and lead time to conjunction in the optimal direction of the deviating actions and the attainable deviation.

The effect of lead time in collision probability is also studied in detail. Uncertainties affecting a possible CA normally reduce as the time of conjunction approaches. Conversely, the larger uncertainties associated to long lead times can hinder the attainable collision probability even if the deviation keeps increasing. To quantify this, minimum collision probability CAMs are designed for scenarios where the covariance matrices of spacecraft and debris are known at the maneuver time. Although this hypothesis differs from the usual practical scenario for spacecraft operators, where the estimated covariance at the CA is provided in the CDM/CSM, the results of this analysis provide a deeper insight on the deviation--uncertainties trade-off for the lead time.

The rest of the manuscript is organized as follows. First, the models for the impulsive CAM between spacecraft and debris, and the State Transition Matrix (STM) relating changes of state at maneuver time with changes of state at CA are presented. Based on these models, the maximum deviation (both total miss distance and impact parameter) and minimum collision probability optimal CAMs are formulated. Then, a sensitivity analysis over conjunction geometry and lead time is performed for the maximum deviation maneuvers. Both approaches, maximum total miss distance and maximum impact parameter in the b-plane, are compared, the evolution of the maneuver orientation is analyzed, and the accuracy of the analytical approximation is assessed. Then, the effect of uncertainties in the CAM is studied by comparing the minimum collision probability and the maximum impact parameter CAMs for a set of cases with uncertainties depending on the lead time. In all cases, b-plane representations of the results allow to gain a better physical insight on the underlying phenomena. Finally, conclusions are drawn.

\section{Dynamical model}\label{sec:DynModel}
Given a CA between an active satellite and a debris, the objective is to perform an impulsive CAM to either maximize the miss distance or minimize the collision probability. The impulsive CAM is performed at a certain time $t_\mathrm{CAM}$ before the time of CA $t_\mathrm{CA}$, with a lead time $\Delta t = t_\mathrm{CA} - t_\mathrm{CAM}$ and an instantaneous change in velocity $\delta \mathbf{v}$. For short $\Delta t$, perturbations other than the impulsive maneuver can be neglected as a first approximation and the two-body problem model is adopted for the motion between $t_\mathrm{CAM}$ and $t_\mathrm{CA}$. The impulsive CAM is modeled through Gauss' planetary equations \cite{vasile2008optimal,battin1999introduction}, giving a linear relation between $\delta \mathbf{v}$ and the instantaneous change in the satellite's Keplerian elements:
\begin{equation}\label{eq:gauss_eq}
\delta \boldsymbol{\alpha} (t_\mathrm{CAM}) = \mathbf{G}_v (t_\mathrm{CAM}) \delta \mathbf{v} (t_\mathrm{CAM}) \, ,
\end{equation}
where  $\delta \boldsymbol{\alpha} = [ \, \delta a \; \delta e \; \delta i \; \delta \Omega \; \delta \omega \; \delta M \, ]^\mathrm{T}$ are the changes in semi-mayor axis, eccentricity, inclination, right ascension of the ascending node, argument of perigee, and mean anomaly, respectively, and $\mathbf{G}_v(t_\mathrm{CAM})$ is the matrix form of the Gauss' planetary equations at $t_\mathrm{CAM}$. The derivation of Gauss' planetary equations can be found in many classical astrodynamics texts (e.g. \cite{battin1999introduction}), and their particular expression for the reference frame consider in this manuscript is reported in Section~\ref{sec:STM_ref_frames} and reference~\cite{vasile2008optimal}. Under the two-body problem model, the modification of the orbital elements at $t_\mathrm{CA}$ coincides with $\delta \boldsymbol{\alpha}(t_\mathrm{CAM})$ except for the mean anomaly, due to the contribution from the change in mean motion. The $\delta M$ at the CA can be written as~\cite{vasile2008optimal}:
\begin{equation}\label{eq:deltaM}
\delta M (t_\text{CA}) = \delta M (t_\text{CAM}) + \delta M_{\delta n} = \delta M (t_\text{CAM}) + \delta n \Delta t \, ,
\end{equation}
where $\delta M_{\delta n}$ represents the change in mean anomaly due to the change in mean motion $\delta n$, which in turn can be related to the change in semi-major axis at the CAM as:
\begin{equation}\label{eq:deltan}
\delta n = \sqrt{\frac{\mu}{a^3}} - \sqrt{ \frac{\mu}{(a+\delta a)^3}} \approx - \frac{3}{2} \frac{\sqrt{\mu}}{a^{5/2}} \delta a \, .
\end{equation}
The evolution of $\delta \boldsymbol{\alpha}$ from $t_\mathrm{CAM}$ to $t_\mathrm{CA}$ is now expressed in matrix form as:
\begin{equation}\label{eq:dalpha_tCAM_2_dalpha_tCA}
\delta \boldsymbol{\alpha}(t_\mathrm{CA}) = \mathbf{G}_M ( \Delta t )  \delta \boldsymbol{\alpha}(t_\mathrm{CAM}) = \left[ \begin{array}{cc}
\mathbf{I}_{5} & \mathbf{0}_{5,1}\\
\begin{array}{ccccc}
-\frac{3}{2} \frac{\sqrt{\mu}}{a^{5/2}} \Delta t  & 0 & 0 & 0 & 0
\end{array} & 1
\end{array} \right]
\delta \boldsymbol{\alpha}(t_\mathrm{CAM}) \, ,
\end{equation}
where $\mathbf{0}_{5,1}$ is the $5 \times 1$ zero matrix and $\mathbf{I}_{5}$ is the $5 \times 5$ identity matrix. $\mathbf{G}_M$ models the change in $\delta\boldsymbol{\alpha}$ during the coasting arc between $t_\mathrm{CAM}$ and $t_\mathrm{CA}$, and its derivation is independent from $\mathbf{G}_v$. Although the previous $\mathbf{G}_M$ only considers the effects due to the CAM, it could be extended to include linearized representations of other perturbations such as $J_2$.

The deviation of the spacecraft at $t_\text{CA}$ is computed analytically from $\delta \boldsymbol{\alpha}$ using linearized relative motion equations~\cite{vasile2008optimal,junkins2009analytical}:
\begin{equation}
\delta \mathbf{r} (t_\mathrm{CA}) = \mathbf{A}_r (t_\mathrm{CA}) \delta \boldsymbol{\alpha} (t_\mathrm{CA}) \, ,
\end{equation}
where $\mathbf{A}_r(t_\mathrm{CA})$ is the matrix form of the linearized relative motion equations. The detailed derivation of $\mathbf{A}_r$ is too long to be reported here and can be found in textbooks such as~\cite{junkins2009analytical}. For convenience, its expression in the reference frame used in this work is reported in Section~\ref{sec:STM_ref_frames}. Plugging in Eqs.~(\ref{eq:gauss_eq},\ref{eq:dalpha_tCAM_2_dalpha_tCA}), a STM $\mathbf{T}$ mapping changes in velocity at $t_\mathrm{CAM}$ with changes in position at $t_\mathrm{CA}$ is reached:
\begin{equation}\label{eq:dr}
\begin{gathered}
\delta \mathbf{r} (t_\mathrm{CA}) =  \mathbf{T} \, \delta \mathbf{v} (t_\mathrm{CAM}) \, ,\\
\mathbf{T} = \mathbf{A}_r (t_\mathrm{CA}) \, \mathbf{G}_M ( \Delta t) \, \mathbf{G}_v (t_\mathrm{CAM}) \, .
\end{gathered}
\end{equation}
In this expression, each matrix models a different contribution to the CAM: $\mathbf{G}_v$ corresponds to the orbit modification due to the impulsive maneuver, $\mathbf{G}_M$ provides the additional change in orbit parameters during the coasting arc, and $\mathbf{A}_r$ maps the orbit modification into a displacement at the CA. This formulation differs slightly from the one used in~\cite{vasile2008optimal,gonzalo2018analysis}, where the contribution to $\delta M$ due to the change in mean motion was incorporated as part of $\mathbf{A}_r$. Although the new approach involves the additional matrix $\mathbf{G}_M$, it is preferred as it clearly separates the three parts of the model (CAM, coasting arc, and displacement evaluation) and more easily allows for the extension to new types of CAMs (such as the low-thrust CAM in~\cite{gonzalo2019introducing,gonzalo2020introducing}) or the inclusion of additional perturbations in the coasting arc.

Equations~(\ref{eq:gauss_eq}-\ref{eq:dr}) are enough for the modeling of impulsive CAMs; however, it would  be convenient to also have the full STM mapping changes in the state at $t_\mathrm{CAM}$ with changes in the state at $t_\mathrm{CA}$. Particularly, this would enable an analytical propagation of covariance matrices instead of requiring computationally expensive methods like Monte-Carlo simulations. The change in Keplerian elements due to a change in state at $t_\mathrm{CAM}$ can be written as:
\begin{equation}
\delta \boldsymbol{\alpha} (t_\mathrm{CAM}) = \left[ \begin{array}{cc}
\mathbf{G}_r (t_\mathrm{CAM}) & \mathbf{G}_v (t_\mathrm{CAM})
\end{array} \right] \delta \mathbf{s} (t_\mathrm{CAM})
= \mathbf{G} (t_\mathrm{CAM}) \, \delta \mathbf{s} (t_\mathrm{CAM}) \, ,
\end{equation}
where $\mathbf{s} = [ \mathbf{r}; \mathbf{v} ]$ is the Cartesian state, $\mathbf{G}_r$ is the partial derivative of $\boldsymbol{\alpha}$ with respect to $\mathbf{r}$ (see the Appendix for a detailed derivation), and $\mathbf{G}_v$ is the matrix form of Gauss' planetary equations previously introduced. Same as before, the change in state at $t_\mathrm{CA}$ is obtained using relative motion equations:
\begin{equation}
\delta \mathbf{s} (t_\mathrm{CA}) = \left[ \begin{array}{c}
\mathbf{A}_r (t_\mathrm{CA}) \\ \mathbf{A}_v (t_\mathrm{CA})
\end{array} \right] \delta \boldsymbol{\alpha} (t_\mathrm{CA}) = \mathbf{A} (t_\mathrm{CA}) \, \delta \boldsymbol{\alpha} (t_\mathrm{CA}) \, .
\end{equation}
Combining both equations together with Eq.~\eqref{eq:dalpha_tCAM_2_dalpha_tCA} for the relation between $\delta \boldsymbol{\alpha}$ at $t_\mathrm{CAM}$ and at $t_\mathrm{CA}$ finally yields:
\begin{equation}\label{eq:STM_full}
\delta \mathbf{s} (t_\mathrm{CA}) = \mathbf{A} (t_\mathrm{CA}) \; \mathbf{G}_M ( \Delta t) \; \mathbf{G} (t_\mathrm{CAM}) \; \delta \mathbf{s} (t_\mathrm{CAM}) = \overline{\mathbf{T}} \; \delta \mathbf{s} (t_\mathrm{CAM}) \, .
\end{equation}
Note that $\overline{\mathbf{T}}$ is a square matrix of dimension $6 \times 6$, and that the reduced STM $\mathbf{T}$ corresponds to the upper-right block of dimension $3 \times 3$ of $\overline{\mathbf{T}}$.

The accuracy of the model will depend on the value of the lead time $\Delta t$. The reason is twofold. On the one hand, by using a two-body problem formulation for the dynamics the effects in time of all orbital perturbations except for the impulsive CAM have been neglected. On the other hand, for a fixed $\delta \mathbf{v}$, as lead time increases so does the deviation at the CA, reducing the accuracy of the linearized relative motion equations. The latter effect is quantified numerically in Section~\ref{sec:max_dev_CAMs}, both for a quasi-circular and an elliptical orbit and several values of $\Delta t$ and $\delta \mathbf{v}$. Regarding the errors due to orbital perturbations, they will strongly depend on the nominal orbit, the physical characteristics of the satellite (e.g. area-to-mass ratio), and the CA configuration. A numerical evaluation of the error between a perturbed and unperturbed model for different configurations can be found in~\cite{bombardelli2015optimal}.

\subsection{STM expression for particular reference frames}\label{sec:STM_ref_frames}
The expressions for $\mathbf{G}$, $\mathbf{G}_M$ and $\mathbf{A}$, and consequently $\mathbf{T}$ and $\overline{\mathbf{T}}$, depend on the reference frames used to project $\delta \mathbf{s}(t_\text{CAM})$ and $\delta \mathbf{s}(t_\text{CA})$. Let us consider two different reference frames: a tangential -- normal -- out-of-plane (TNH) frame at $t_\text{CAM}$ and a radial -- transversal -- out-of-plane (RTH) frame at $t_\text{CA}$. The TNH frame $\mathcal{T}=\left\{ S; \hat{t} , \hat{n}, \hat{h} \right\}$ is centered at the spacecraft's position and its axes are given by the tangential direction (i.e. along the velocity), the normal direction (inward belonging to the orbital plane), and the perpendicular-to-the-orbit-plane direction. The unit vectors for the TNH frame can be calculated as:
\begin{equation}
\mathbf{i}_{\hat{t}} = \frac{\mathbf{v}}{||\mathbf{v}||} \, , \qquad \mathbf{i}_{\hat{h}} = \frac{ \mathbf{r}\times \mathbf{v}}{||\mathbf{r}\times \mathbf{v}||} \, , \qquad \mathbf{i}_{\hat{n}} = \mathbf{i}_{\hat{h}} \times \mathbf{i}_{\hat{t}} \, .
\end{equation}
where $\mathbf{r}$ and $\mathbf{v}$ are the inertial position and velocity. The RTH frame $\mathcal{R}=\left\{ S; \hat{r} , \hat{\vartheta}, \hat{h} \right\}$ is also centered at the spacecraft, and its axes are oriented along the radial, transversal, and perpendicular-to-the-orbit-plane directions, respectively. In mathematical form:
\begin{equation}
\mathbf{i}_{\hat{r}} = \frac{\mathbf{r}}{||\mathbf{r}||} \, , \qquad \mathbf{i}_{\hat{h}} = \frac{ \mathbf{r}\times \mathbf{v}}{||\mathbf{r}\times \mathbf{v}||} \, , \qquad \mathbf{i}_{\hat{\vartheta}} = \mathbf{i}_{\hat{h}} \times \mathbf{i}_{\hat{r}} \, .
\end{equation}
The expression for $\mathbf{G}_v$, with $\delta \mathbf{v}$ projected in the TNH frame, can be found in the literature~\cite{vasile2008optimal,battin1999introduction}:
\begin{equation}
\mathbf{G}_v (t_\text{CAM}) = \left[ \begin{array}{ccc}
\frac{2 a^2 v}{\mu} & 0 & 0 \\
\frac{2(e+\cos f)}{v} & - \frac{r}{a v}\sin f & 0 \\
0 & 0 & \frac{r \cos\theta}{h} \\
0 & 0 & \frac{r \sin\theta}{h \sin i} \\
\frac{2 \sin f}{e v} & \frac{2e + (r/a)\cos f}{e v} & -\frac{r \sin\theta \cos i}{h \sin i}\\
-\frac{b}{e a v} 2 \left( 1+ \frac{e^2 r}{p}\right) \sin f & -\frac{b}{e a v}\frac{r}{a}\cos f & 0
\end{array} \right] \, ,
\end{equation}
where $f$ is the true anomaly, $\theta=f+\omega$ is the argument of latitude, $b$ is the semi-minor axis of the elliptic orbit, $p$ is its parameter, $h$ is the norm of the angular momentum, $v$ is the magnitude of the velocity, and $r$ is the radial distance. The expression for $\mathbf{G}_r$ with $\delta\mathbf{r}$ projected in a generic reference frame is developed in the Appendix; particularizing for the TNH frame by setting $\mathbf{r}=[ \; r_t \; r_n \; 0 \; ]^\top$ and $\mathbf{v}=[ \; v \; 0 \; 0 \; ]^\top$ one reaches:  
\begin{equation}
\mathbf{G}_r (t_\text{CAM}) = \left[ \begin{array}{ccc}
\frac{2 a^2}{r^3} r_t & \frac{2 a^2}{r^3} r_n & 0 \\
\frac{1}{\mu a e} \left( \frac{h^2 a}{r^3}-v^2 \right) r_t + \frac{r \sin f}{a h} v & \frac{1}{\mu a e} \left( \frac{h^2 a}{r^3}-v^2 \right) r_n & 0 \\
0 & 0 & \frac{\sin\theta+e\sin\omega}{p} \\
0 & 0 & -\frac{\cos\theta+e\cos\omega}{p \sin i} \\
G_{51} r_t + G_{52} v & G_{51} r_n & G_{53} \\
G_{61} r_t + G_{62} v & G_{61} r_n & 0
\end{array} \right] \, ,
\end{equation}
with
\begin{gather*}
G_{51} = -\frac{r}{h^2 e}\sin f \left(\frac{h^2}{p r^3} \left(p+e^2 r \right)-\frac{(p+r) v^2}{r^2} \right) \, , \\
G_{52} = -\frac{r}{h e p} \left(\cos f +e \right)\, , \qquad G_{53} = \left( \cos\theta+e \cos\omega \right) \frac{\cos i}{p \sin i} \, , \\
G_{61} = \frac{b}{a^2 e p r^2} \left( r^2 - a (p+r)\right) \sin f \, , \qquad G_{62} = \frac{r b}{h a^2 e} \cos f \, ,
\end{gather*}
where $r_t$ and $r_n$ are the projections of the position vector along the tangential and normal directions, respectively.

Similarly, the equation corresponding to $\mathbf{A}_r$, with $\delta \mathbf{r}$ in RTH frame, can be taken from Schaub and Junkins~\cite{junkins2009analytical}:
\begin{equation}
\mathbf{A}_r^\text{T} (t_\text{CA}) = \left[ \begin{array}{ccc}
\frac{r}{a} & 0 & 0 \\
-a \cos f & \frac{r \sin f }{\gamma^2} \left( 2 + e \cos f \right) & 0 \\
0 & 0 & r \sin \theta \\
0 & r \cos i & - r \cos \theta \sin i \\
0 & r & 0 \\
\frac{a e \sin f }{\gamma} & \frac{r}{\gamma^3} \left( 1 + e \cos f \right)^2 & 0
\end{array} \right] \, ,
\end{equation}
where $\gamma=\sqrt{1-e^2}$. Regarding the relative velocity due to $\delta \boldsymbol{\alpha}$, Schaub and Junkins~\cite{junkins2009analytical} provide the full, non-linear expressions for $\delta \mathbf{v}$ with respect to the non-inertial rotating frame of the relative motion, which in this case is centered at the nominal position of the CA. Linearizing these expressions and adding the terms due to the rotation of the frame, an expression for $\mathbf{A}_v$ with $\delta \mathbf{v}$ in RTH frame is reached:
\begin{equation}
\mathbf{A}_v^\text{T} (t_\text{CA}) = \left[ \begin{array}{ccc}
-\frac{e h \sin f }{2 a p} & -\frac{h}{2 a r} & 0 \\
-\frac{ h \sin f }{\gamma^2 r} & \frac{h(e+\cos f)}{\gamma^2 p} & 0 \\
0 & 0 & \frac{h}{p} \left(e\cos\omega+\cos\theta \right) \\
- \frac{h \cos i}{r} & \frac{e h}{p} \cos i \sin f & \frac{h \sin i}{p} \left(e\sin\omega+\sin\theta\right) \\
- \frac{h}{r} & \frac{e h}{p} \sin f & 0 \\
-\frac{a h}{\gamma r^2} & 0 & 0
\end{array} \right] \, .
\end{equation}

\subsection{B-plane projection}
A more convenient representation of the spacecraft's deviation can be achieved using the b-plane~\cite{opik1976interplanetary,kizner1961method}, defined through a local reference frame $\mathcal{B}=\left\{ D; \xi , \eta, \zeta \right\}$ centered at the debris $D$ and with unit vectors:
\begin{equation}
\hat{\boldsymbol\eta} = \frac{ \mathbf{v}_{SC} - \mathbf{v}_D }{|| \mathbf{v}_{SC} - \mathbf{v}_D ||} , \qquad \hat{\boldsymbol\xi} = \frac{ \mathbf{v}_D \times \hat{\boldsymbol\eta} }{|| \mathbf{v}_D \times \hat{\boldsymbol\eta} ||}, \qquad \hat{\boldsymbol\zeta} = \hat{\boldsymbol\xi} \times \hat{\boldsymbol\eta} \, ,
\end{equation}
where $\mathbf{v}_D$ and $\mathbf{v}_{SC}$ are the velocities of debris and spacecraft, respectively. The b-plane of the encounter is then the plane $\zeta-\xi$, orthogonal to the relative velocity of the spacecraft with respect to the debris. Moreover, it can be checked that the $\zeta$ axis is oriented along the direction opposite to the projection of $\mathbf{v}_D$ onto the b-plane. One of the main advantages of using the b-plane for the design of CAMs is that displacements in the $\zeta$ axis are associated with phasing maneuvers (i.e., time shift), whereas displacements in the $\xi$ axis come from a geometrical change in the spacecraft's orbit to modify the Minimum Orbit Intersection Distance (MOID). For this reason, from now on $\zeta$ will be referred to as the time axis, and $\xi$ will be called geometry axis.

In the classic b-plane theory the impact parameter $b^*$ is defined as the intersection of the incoming asymptote of the relative hyperbolic trajectory and the b-plane. Assuming that the trajectory of the spacecraft with respect to the debris at the CA is nearly rectilinear, a $b^*$ can be defined for our case in an analogous way (although the relative trajectory is not hyperbolic), and it will be a good approximation of the minimum miss distance~\cite{vasile2008optimal}. The condition of nearly-rectilinear relative trajectory will be fulfilled if the duration of the CA is small
compared with the orbital period of the objects (short-term encounter). Furthermore, $b^*$ will also be close to the actual intersection of the spacecraft's trajectory with the b-plane. Because of this, from now on the distance between the debris and the spacecraft on the b-plane will also be referred to as impact parameter.

The b-plane projection of $\delta \mathbf{r}$ can be expressed in matrix form as~\cite{petit2018optimal}:
\begin{equation}
\delta \mathbf{b}^* = \mathbf{M}_{\delta \mathbf{b}^*} \delta \mathbf{r} \, ,
\end{equation}
with
\begin{equation}
\mathbf{M}_{\delta \mathbf{b}^*} =\left[ \begin{array}{ccc}
\hat{\eta}_2^2 + \hat{\eta}_3^2 & - \hat{\eta}_1 \hat{\eta}_2 & - \hat{\eta}_1 \hat{\eta}_3 \\
- \hat{\eta}_1 \hat{\eta}_2 & \hat{\eta}_1^2 + \hat{\eta}_3^3 & - \hat{\eta}_2 \hat{\eta}_3 \\
- \hat{\eta}_1 \hat{\eta}_3 & - \hat{\eta}_2 \hat{\eta}_3 & \hat{\eta}_1^2 + \hat{\eta}_2^2
\end{array} \right] \, ,
\end{equation}
where $\hat{\eta}_1$, $\hat{\eta}_2$, and $\hat{\eta}_3$ are the components of unit vector $\hat{\boldsymbol\eta}$, expressed in the same reference frame as $\delta \mathbf{r}$. Recalling Eq.~\eqref{eq:dr}, the deviation in the b-plane for a given $\delta \mathbf{v}$ is:
\begin{equation}\label{eq:db}
\begin{gathered}
\delta \mathbf{b}^* = \mathbf{Z} \, \delta \mathbf{v} \, ,\\
\mathbf{Z} = \mathbf{M}_{\delta \mathbf{b}^*} \, \mathbf{T} \, .
\end{gathered}
\end{equation}

\subsection{Maximum miss distance CAM}
The mathematical formulation for the maximum miss distance CAM with $\delta v \leq \delta v_\mathrm{max}$ is now presented. Assuming a direct impact (zero miss distance at nominal CA), the objective function for maximum deviation in terms of $\delta \mathbf{r}$, $J_{\delta r}$, can be written as:
\begin{equation}
J_{\delta r} = || \delta \mathbf{r} || = || \mathbf{T} \delta \mathbf{v} || = \delta \mathbf{v}^\text{T} \, \mathbf{T}^\text{T} \mathbf{T} \, \delta \mathbf{v} \, .
\end{equation}
Analogously, the objective function for maximum impact parameter, $J_{\delta b^*}$, takes the form:
\begin{equation}
J_{\delta b^*} = || \delta \mathbf{b}^* || = || \mathbf{Z} \delta \mathbf{v} || = \delta \mathbf{v}^\text{T} \, \mathbf{Z}^\text{T} \mathbf{Z} \, \delta \mathbf{v} \, .
\end{equation}
In both cases, the impulsive CAM has to fulfill the constraint $\delta v \leq \delta v_\mathrm{max}$. This constraint is important not only for operational reasons, but also because both objective functions are unbounded (that is, they would lead to an infinite displacement if no $\delta v$ constraint is applied).

We can use these expressions to proceed in finding the optimal maneuver given at a certain time to maximize the miss distance at CA. Following the approach proposed by Conway~\cite{conway2001near}, maximizing $J_{\delta r}$ is equivalent to maximizing the associated quadratic form by choosing a $\delta \mathbf{v}_\text{opt}$ parallel to the eigenvector of $\mathbf{T}^\text{T} \mathbf{T}$ conjugated to the maximum eigenvalue. Note that the sign of $\delta \mathbf{v}_\text{opt}$ is not defined as it does not affect the magnitude of the deviation~\cite{vasile2008optimal,conway2001near}. Regarding the magnitude of the impulse $\delta v_\text{opt}$, it is straightforward to check that the quadratic form is maximized by using all the available impulse capability $\delta v_\mathrm{max}$. The same procedure can be applied for the maximization of $J_{\delta b^*}$ by solving the eigenvalue problem for $\mathbf{Z}^\text{T} \mathbf{Z}$.

\subsection{Minimum collision probability CAM}
Owing to the uncertainties in the orbital state of spacecraft and debris, the maximum deviation CAM may differ significantly from the minimum collision probability one. In a recent work, Bombardelli and Hernando-Ayuso~\cite{bombardelli2015optimal} proposed an approximate analytical method for designing minimum collision probability CAMs, using Chan's approach for the computation of collision probabilities~\cite{chan2008spacecraft,chan2009international} in order to reduce the optimization problem to a quadratic form similar to the one considered by Conway for maximum deflection~\cite{conway2001near}. Following Chan's method, the original conjunction in the b-plane, in which each object has its own spherical envelope and covariance matrix, is reduced to an equivalent problem by assigning a combined covariance to the debris (with no envelope) and a combined envelope to the spacecraft (with no covariance). The spherical envelope for each object in the original problem is a sphere encompassing the whole object and centered at its center of mass, whereas the combined envelope is centered at the spacecraft and its radius $r_A$ is equal to the sum of the radii of the individual envelopes. The combined covariance in the b-plane reference frame:
\begin{equation}
\mathbf{C} = \left[ \begin{array}{cc}
\sigma_\xi^2 & \rho_{\xi\zeta} \sigma_\xi \sigma_\zeta \\
\rho_{\xi\zeta} \sigma_\xi \sigma_\zeta & \sigma_\zeta^2
\end{array} \right] \, ,
\end{equation}
can be calculated as the sum of the individual covariances for both objects, provided that their determination is statistically independent. Then, the collision probability between debris and spacecraft can be approximated through the convergent series:
\begin{equation}
P(u,w) = \mathrm{e}^{-w/2} \sum_{m=0}^\infty \frac{w^m}{2^m m!} \left( 1 - \mathrm{e}^{-u/2} \sum_{k=0}^m \frac{u^k}{2^k k!} \right) \, ,
\end{equation}
with
\begin{equation}
u = \frac{r_A^2}{\sigma_\xi \sigma_\zeta \sqrt{1-\rho_{\xi\zeta}^2}} \, ,
\end{equation}
\begin{equation}
w = \left[ \left( \frac{\xi}{\sigma_\xi} \right)^2 + \left( \frac{\zeta}{\sigma_\zeta} \right)^2 - 2 \rho_{\xi\zeta} \frac{\xi}{\sigma_\xi} \frac{\zeta}{\sigma_\zeta} \right]/(1-\rho_{\xi\zeta}^2) \, ,
\end{equation}
where $(\xi,\zeta)$ is the position of the spacecraft in the b-plane at the CA. As shown by Chan~\cite{chan2008spacecraft,chan2009international}, accurate results can be obtained for $m=3$ for small values of $u$. Interestingly, the position of the spacecraft only influences collision probability $P$ through the depth of intrusion $w$. Then, Bombardelli and Hernando-Ayuso~\cite{bombardelli2015optimal} prove that minimizing $P$ is equivalent to maximizing an objective function $J_P$ defined as:
\begin{equation}
J_P = \left( \frac{\xi}{\sigma_\xi} \right)^2 + \left( \frac{\zeta}{\sigma_\zeta} \right)^2 - 2 \rho_{\xi\zeta} \frac{\xi}{\sigma_\xi} \frac{\zeta}{\sigma_\zeta} \, .
\end{equation}
$J_P$ can be written in matrix form as:
\begin{equation}
J_P = \delta \mathbf{r}^\text{T} \, \mathbf{Q}^* \, \delta \mathbf{r} = {\delta \mathbf{b}^*}^\text{T} \, \mathbf{Q}^* \, \delta \mathbf{b}^* \, ,
\end{equation}
where $\delta \mathbf{r}$ and $\delta \mathbf{b}^*$ are expressed in the b-plane reference frame, and:
\begin{equation}
\mathbf{Q}^* = \left[ \begin{array}{ccc}
1/\sigma_{\xi}^2 & 0 & -\rho_{\xi\zeta} / \sigma_{\xi}\sigma_{\zeta} \\
0 & 0 & 0 \\
-\rho_{\xi\zeta} / \sigma_{\xi}\sigma_{\zeta} & 0 & 1/\sigma_\zeta^2
\end{array} \right] \, ,
\end{equation}
also projected in the b-plane reference frame. By substituting the expressions for $\delta \mathbf{r}$ or $\delta \mathbf{b}^*$, Eqs.~\eqref{eq:dr} and \eqref{eq:db}, the problem is reduced to the maximization of a quadratic form of $\delta \mathbf{v}$, which can be solved in the same fashion as the maximum deviation case. The main difference with respect to the work by Bombardelli and Hernando-Ayuso~\cite{bombardelli2015optimal} is the choice of dynamical model: while they develop a STM based on the Dromo orbital elements (for which time is a dependent variable), we apply the previously introduced STM based on Gauss' planetary equations and linear relative motion. This approach has the advantage of not requiring the solution of the time equation for every variation of the orbit of the spacecraft, reducing computational cost as it is fully analytical, while the solution of the time equation requires a numerical solver. Conversely, the formulation by Bombardelli and Hernando-Ayuso~\cite{bombardelli2015optimal} can be more convenient for geometry-based studies, as the independent variable is related to the true anomaly. Furthermore, having time as a dependent variable may lead to better accuracy for long $\Delta t$. Quantifying these effects would require a comparative analysis that lies out of the scope of this manuscript.

\section{CAM Design and Sensitivity Analysis}
This section deals with the design of maximum deviation and minimum collision probability CAMs between a spacecraft and a space debris in different practical scenarios, using the models presented in Section~\ref{sec:DynModel}. Realistic test data for the CA is generated considering current missions for the maneuvering spacecraft, and statistical data from ESA MASTER-2009 tool~\cite{flegel2009master} for the debris. The lack of information about the corresponding covariance matrices is tackled by considering a reference covariance constructed from TLEs and a corrective procedure to account for the change in true anomaly. Maximum deviation CAMs are studied first, performing an extensive sensitivity analysis over the geometry of the conjunction. The advantages of working in the b-plane are explored, and the accuracy of the analytical solutions is assessed. Then, the effect of uncertainties is studied by considering a nominal CA and exploring the behavior of the minimum collision probability CAM with its lead time for different evolutions of the covariance matrices.

\subsection{Maximum deviation CAMs}\label{sec:max_dev_CAMs}
The models presented in Section~\ref{sec:DynModel} encompass two different approaches for maximum deviation CAMs: maximizing either the total miss distance $\delta r$ or the impact parameter in the b-plane $\delta b^*$. An extensive sensitivity analysis on the geometry of the CA is now performed comparing both approaches.

Two test cases are selected from current ESA missions: PROBA-2 (quasi-circular orbit) and the XXM-Newton observatory (highly elliptical orbit). Their nominal orbital parameters considered for this study are reported in Table~\ref{tab:proba2_xmm}\footnote{Orbit information retrieved from \url{http://www.heavens-above.com}  [Last retrieved 30/04/2018]}. For the study, the nominal orbit geometry will be preserved (i.e semi-major axis, eccentricity, inclination, right ascension of the ascending node, and argument of perigee), whereas the position inside the orbit (i.e. the true anomaly) will be changed to study its influence in the CAM.
\begin{table}
\centering
\caption{Nominal orbital parameters for PROBA-2 and XMM}\label{tab:proba2_xmm}
\begin{tabular}{cccccccc}
\hline\hline 
Object & Epoch [UTC] & $a$ [km] & $e$ [-] & $i$ [deg] & $\Omega$ [deg] & $\omega$ [deg] & $M$ [deg] \\ 
\hline 
PROBA-2 & 2018/04/20 03:18:34 & 7093.637 & 0.0014624 & 98.2443 & 303.5949 & 109.4990 & 250.7787 \\
XXM & 2018/04/27 18:31:05 & 66926.137 & 0.8031489 & 70.1138 & 348.8689 & 95.9905 & 359.6770 \\ 
\hline\hline
\end{tabular}
\end{table}

\begin{figure}[t]
	\centering
	\includegraphics[width=0.47\textwidth]{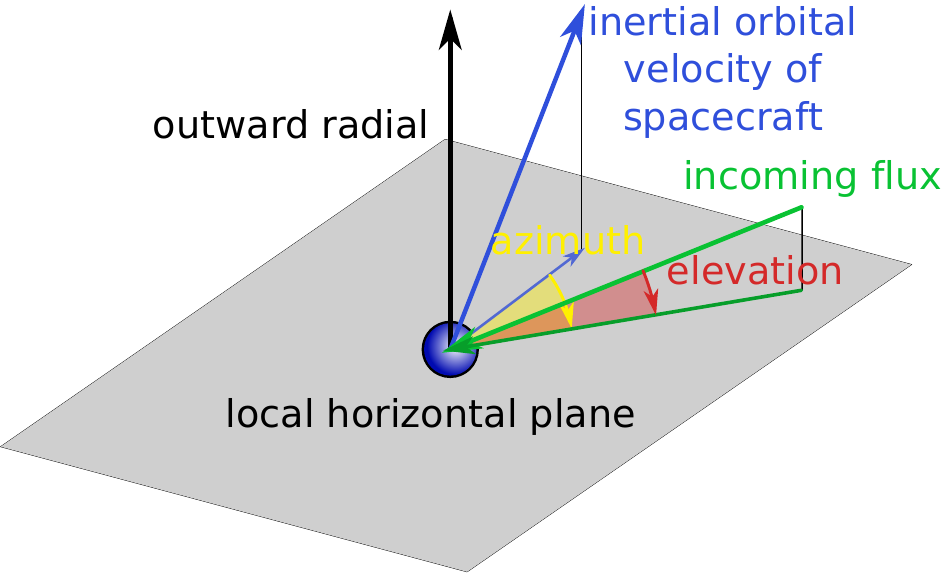}
	\caption{Conjunction geometry parametrization in ESA MASTER-2009.}\label{fig:Diagram_MASTER}
\end{figure}

\begin{figure}
	\centering
	\includegraphics[width=0.49\textwidth]{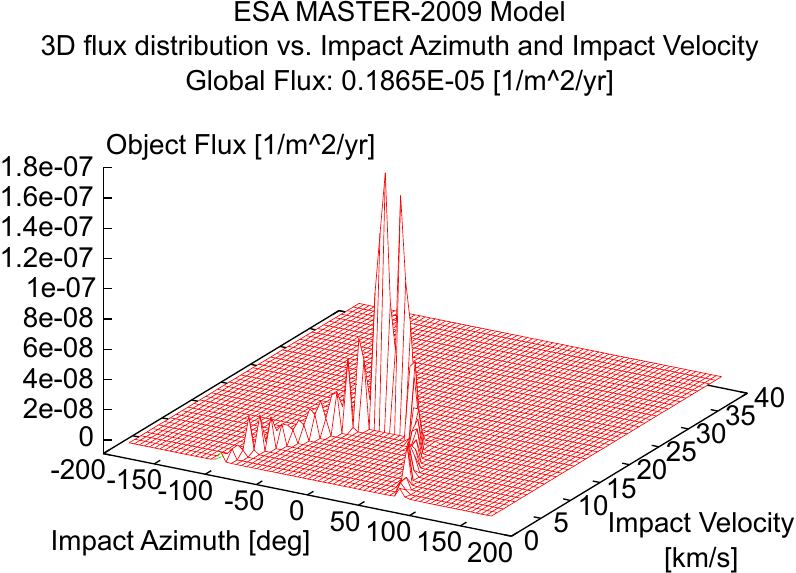}\hfill
	\includegraphics[width=0.49\textwidth]{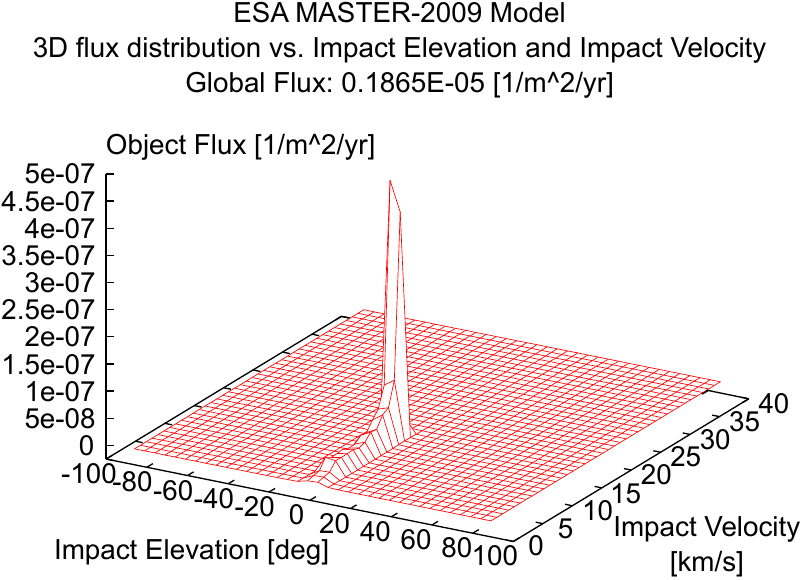}
	\caption{Debris object fluxes for PROBA-2 nominal orbit from ESA MASTER-2009, as functions of azimuth, elevation and relative velocity at conjunction.}\label{fig:PROBA2_MASTER}
\end{figure}

\begin{figure}
	\centering
	\includegraphics[width=0.49\textwidth]{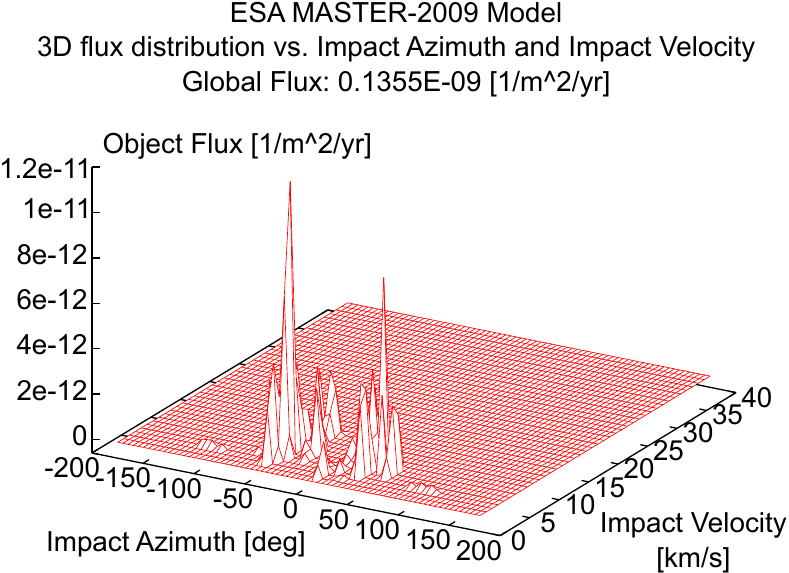}\hfill
	\includegraphics[width=0.49\textwidth]{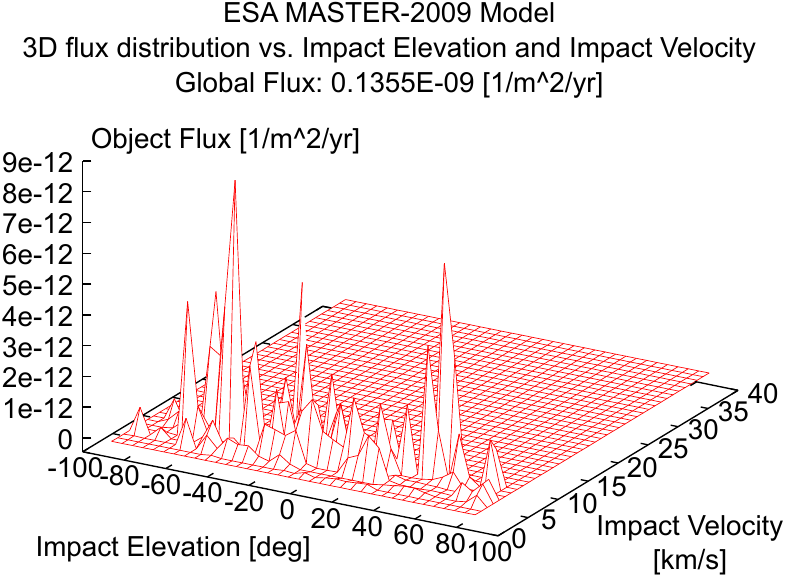}
	\caption{Debris object fluxes for XMM nominal orbit from ESA MASTER-2009, as functions of azimuth, elevation and relative velocity at conjunction.}\label{fig:XMM_MASTER}
\end{figure}

Regarding the debris, statistical information from MASTER-2009~\cite{flegel2009master} is used to cover a wide range of realistic conjunction geometries. For each position of the spacecraft in its nominal orbit, a set of possible debris orbits with zero miss distance at CA (direct impact) is constructed from the ranges in conjunction azimuth, elevation and relative velocity given by MASTER. In all cases, the sources for conjunctions are launchers and mission related objects. Figure~\ref{fig:Diagram_MASTER} shows the definition of the conjunction azimuth and elevation angles, whereas the object flux distributions as functions of azimuth, elevation and relative velocity for PROBA-2 and XMM are given in Figs.~\ref{fig:PROBA2_MASTER} and \ref{fig:XMM_MASTER}, respectively. Clear patterns can be appreciated for the PROBA-2 case in Fig.~\ref{fig:PROBA2_MASTER}, yielding compact ranges for the conjunction geometry and relative velocity. Particularly, significant collision probabilities are only obtained for elevations close to 0 deg. Furthermore, not all combinations inside the outer bounding values correspond to a high object flux, as clearly appreciated in the plot for impact azimuth and relative velocity. The object flux distributions for the XMM nominal orbit in Fig.~\ref{fig:XMM_MASTER} show a less regular behavior, but it is still straightforward to define bounds for the conjunction.

Because the magnitude of the impulsive maneuver does not affect its optimal direction in the linearized model, there are four free parameters to be considered in the sensitivity analysis: true anomaly $f_0$ of the spacecraft at the conjunction, relative velocity $\Delta V$ between the spacecraft and the debris at the conjunction, and azimuth and elevation angles of the debris in the encounter frame (geometry at the conjunction). To present the data in a concise manner, the last two are removed by presenting only the case with the maximum deviation (that is, the one with the most effective CAM) for each combination of true anomaly and relative velocity.

\begin{figure}
	\centering
	\includegraphics[width=0.60\textwidth]{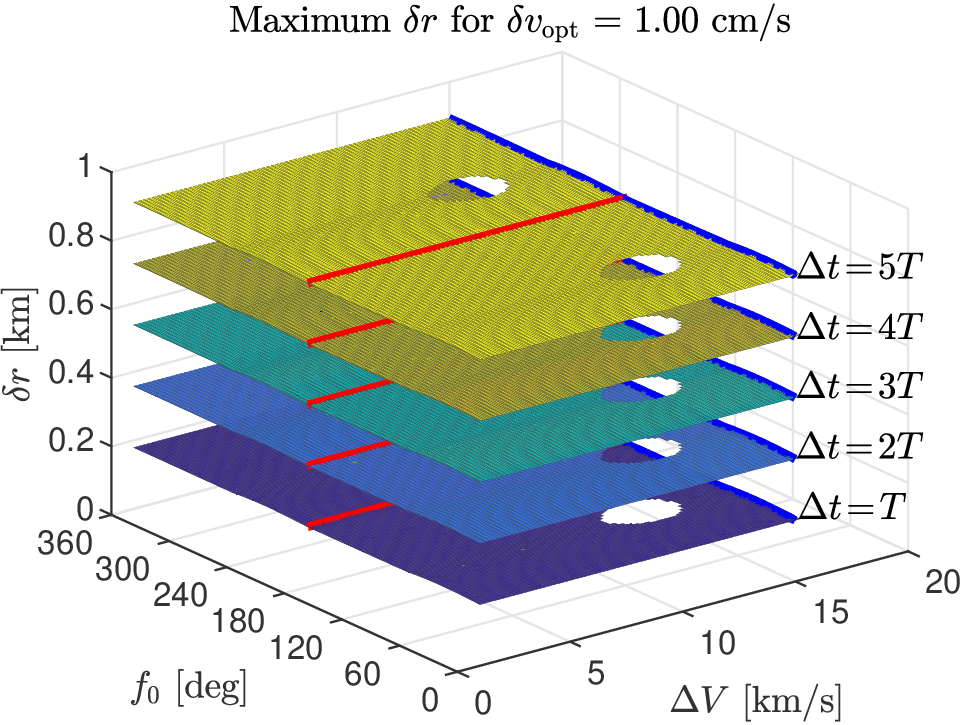}
	\caption{Maximum miss distance for PROBA-2 for an impulsive CAM of  1~cm/s.}\label{fig:PROBA2_dr_surf}
\end{figure}

\begin{figure}
	\centering
	\includegraphics[width=0.49\textwidth]{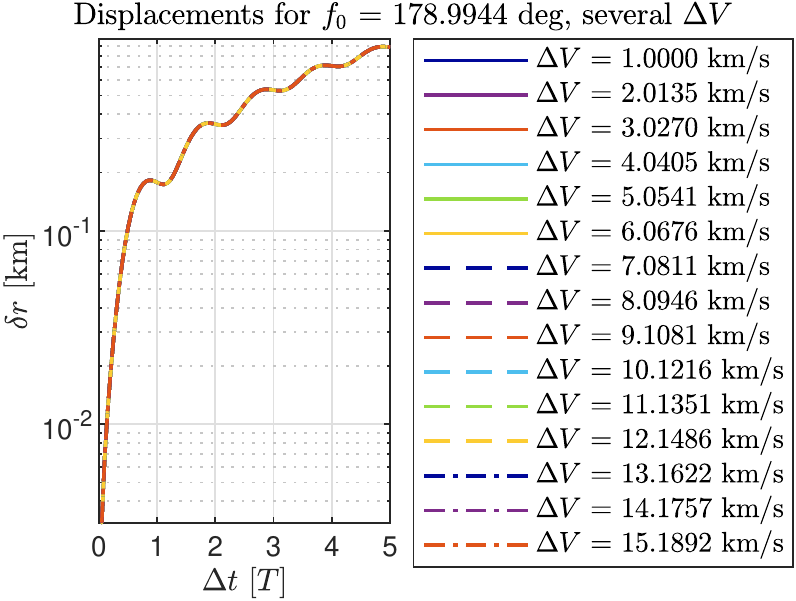}\hfill
	\includegraphics[width=0.49\textwidth]{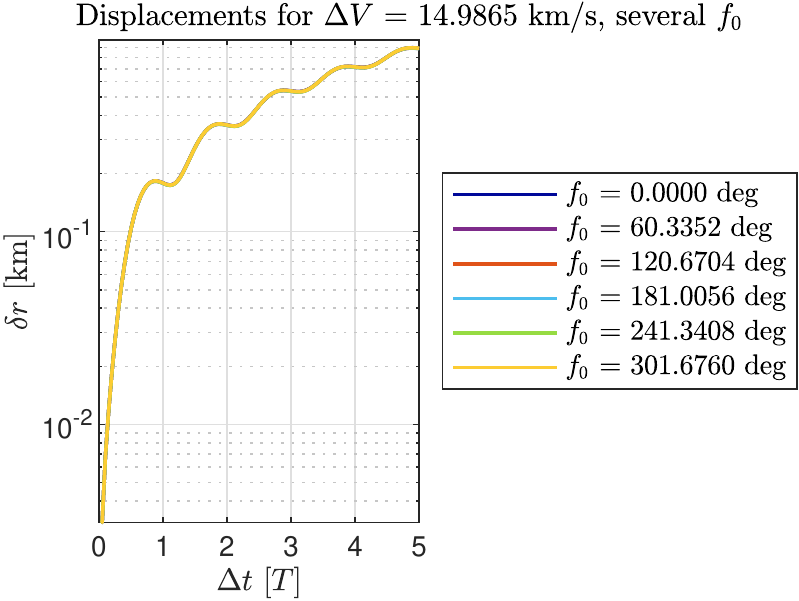}
	\caption{Maximum miss distance for PROBA-2, for a CAM of 1~cm/s and several values of $f_0$ and $\Delta V$.}\label{fig:PROBA2_drmax}
\end{figure}

The maximum miss distance CAM results for PROBA-2 are presented in Figs.~\ref{fig:PROBA2_dr_surf}-\ref{fig:PROBA2_drmax}, for a $\delta v_\text{opt}$ of $1$~cm/s. Because $\delta r$ depends linearly with $\delta v$ in the analytical formulation, results for a different $\delta v_\mathrm{opt}$ can be obtained by scaling. Figure~\ref{fig:PROBA2_dr_surf} summarizes the effect of true anomaly, relative velocity at conjunction, and maneuver lead time in the attainable $\delta r$. It is straightforward to check that the former two have a very small effect on the miss distance, whereas the lead time increases $\delta r$ in a regular fashion. Note that, for some combinations of $f_0$ and $\Delta V$ no feasible conjunction was found, in the sense that all resulting candidate debris orbits were hyperbolic, leading to holes in the constant-lead-time surfaces. In other words, for these combinations of true anomaly and relative velocity, all of the azimuth and elevation angles considered in the range given by MASTER led to hyperbolic orbits, which cannot be the case for an Earth-bound debris and were discarded. This is only related to the creation of the synthetic debris population, and has no implication on the applicability of the maximum deviation CAM model in those regions. Going into further detail, Fig.~\ref{fig:PROBA2_drmax} represents the evolution of $\delta r$ with lead time for fixed values of $f_0$ and $\Delta V$. For clarity, these fixed values have been marked also in Fig.~\ref{fig:PROBA2_dr_surf}, with a red line parallel to the $\Delta V$ axis for fixed $f_0$, and a blue line parallel to the $f_0$ axis for fixed $\Delta V$. As previously indicated, true anomaly and relative conjunction velocity have a negligible effect on the miss distance.

\begin{figure}
	\centering
	\includegraphics[width=0.60\textwidth]{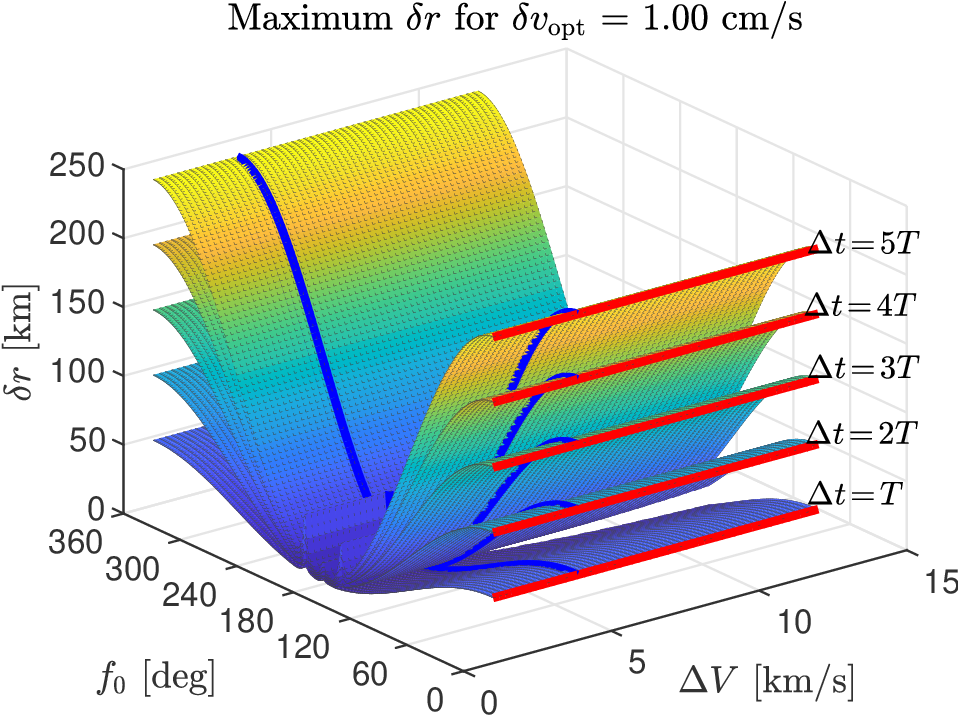}
	\caption{Maximum miss distance for XMM for an impulsive CAM of 1~cm/s.}\label{fig:XMM_dr_surf}
\end{figure}

\begin{figure}
	\centering
	\includegraphics[width=0.49\textwidth]{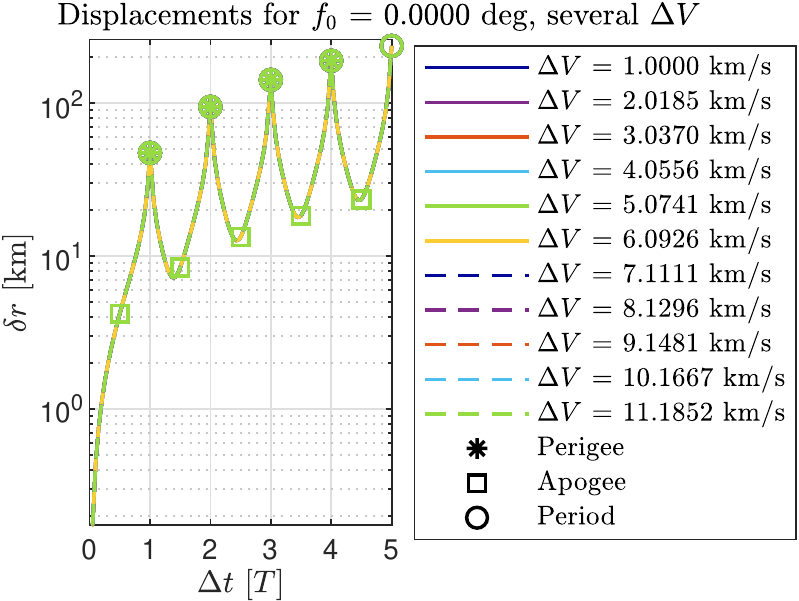}\hfill
	\includegraphics[width=0.49\textwidth]{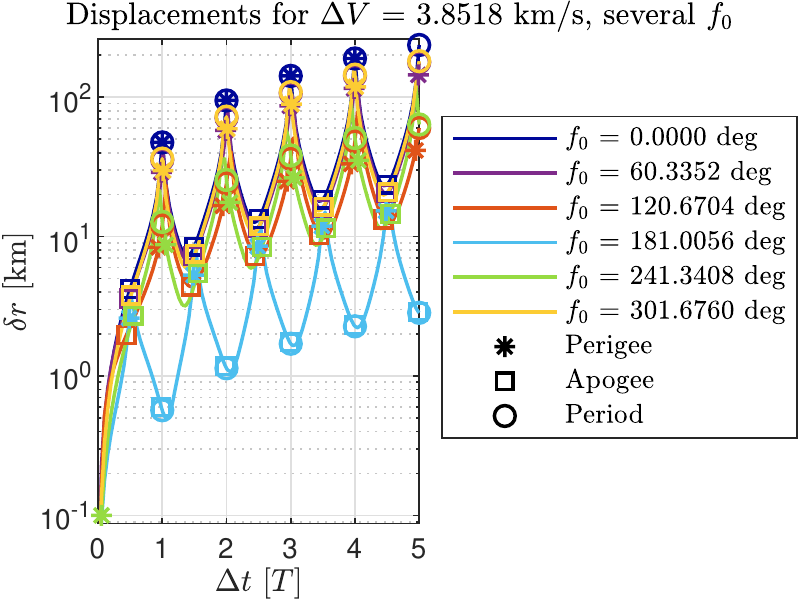}
	\caption{Maximum miss distance for XMM, for a CAM of 1~cm/s and several values of $f_0$ and $\Delta V$.}\label{fig:XMM_drmax}
\end{figure}

\begin{figure}
	\centering
	\includegraphics[width=0.49\textwidth]{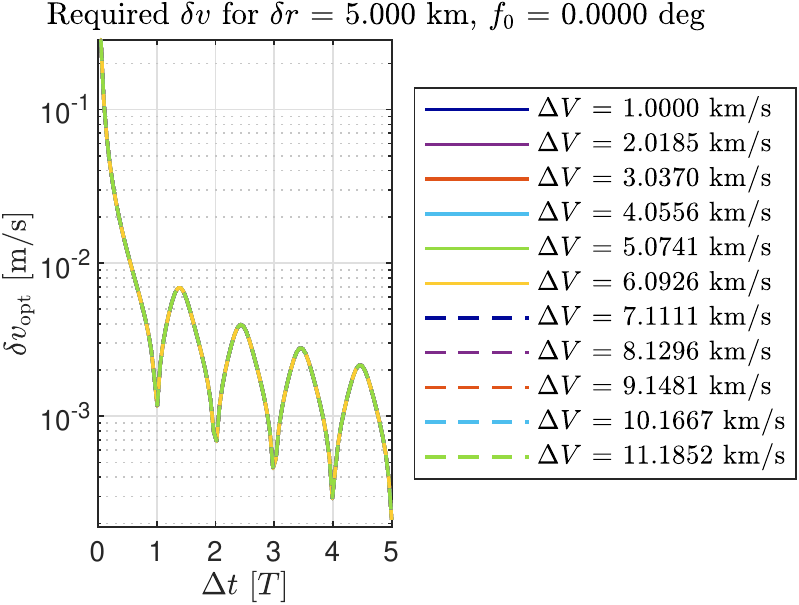}\hfill
	\includegraphics[width=0.49\textwidth]{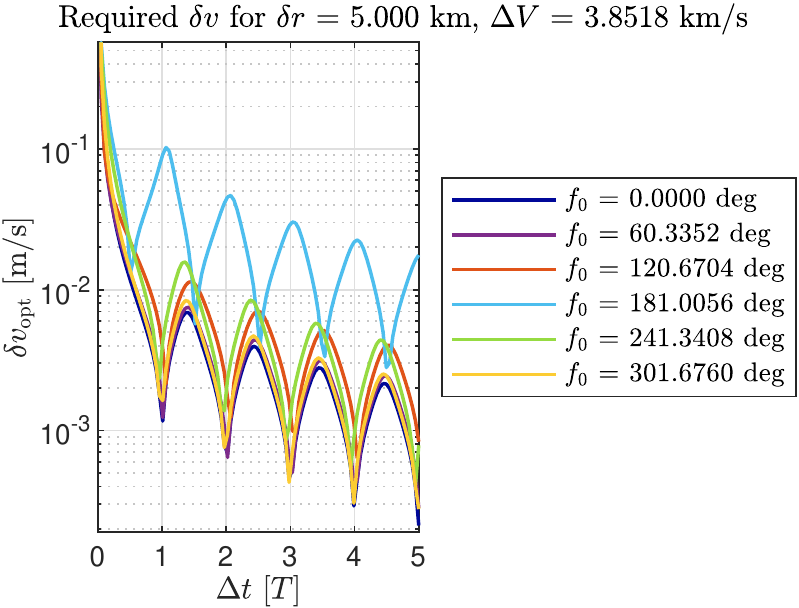}
	\caption{Required $\delta v_\text{opt}$ to reach a miss distance of 5~km for XMM, for several values of $f_0$ and $\Delta V$.}\label{fig:XMM_dvreq_5km}
\end{figure}

The effect of the orbital eccentricity of the spacecraft can be clearly appreciated in Figs.~\ref{fig:XMM_dr_surf}-\ref{fig:XMM_dvreq_5km} for the XMM test case. While the variation of $\delta r$ with $\Delta V$ is still negligible, it now strongly depends on the position of the spacecraft inside its orbit at the CA. Same as before, the fixed $f_0$ and fixed $\Delta V$ plots in Fig.~\ref{fig:XMM_drmax} correspond, respectively, to the red and blue lines in Fig.~\ref{fig:XMM_dr_surf}. Regarding the results for fixed $\Delta V$, it is important to highlight how the best solution for each period corresponds to performing the maneuver close to the perigee, and that the highest $\delta r$ is reached when the maneuver also coincides with a whole number of periods (i.e. when the CA takes place at the perigee). Conversely, minimum values for $\delta r$ are obtained close to the apogees, and results worsen as the true anomaly of the CA separates from the perigee of the maneuvering spacecraft's orbit. Analogous behaviors are reproduced for the required $\delta v_\text{opt}$ for a fixed $\delta r$ of $5~\mathrm{km}$, represented in Fig.~\ref{fig:XMM_dvreq_5km}. Note that, because the formulation is linear in $\delta v_\text{opt}$, Fig.~\ref{fig:XMM_dvreq_5km} replicates the evolution of  Fig.~\ref{fig:XMM_drmax} with the vertical axis reversed.

\begin{figure}
	\centering
	\includegraphics[width=0.60\textwidth]{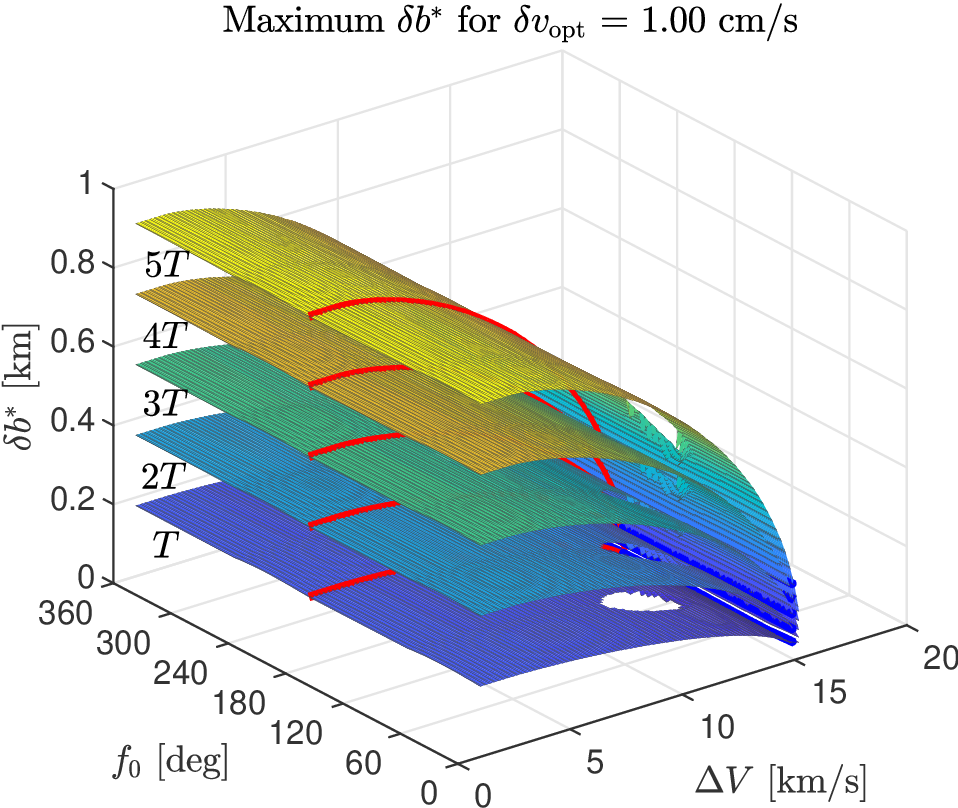}
	\caption{Maximum impact parameter for PROBA-2 for an impulsive maneuver of  1~cm/s.}\label{fig:PROBA2_db_surf}
\end{figure}

\begin{figure}
	\centering
	\includegraphics[width=0.49\textwidth]{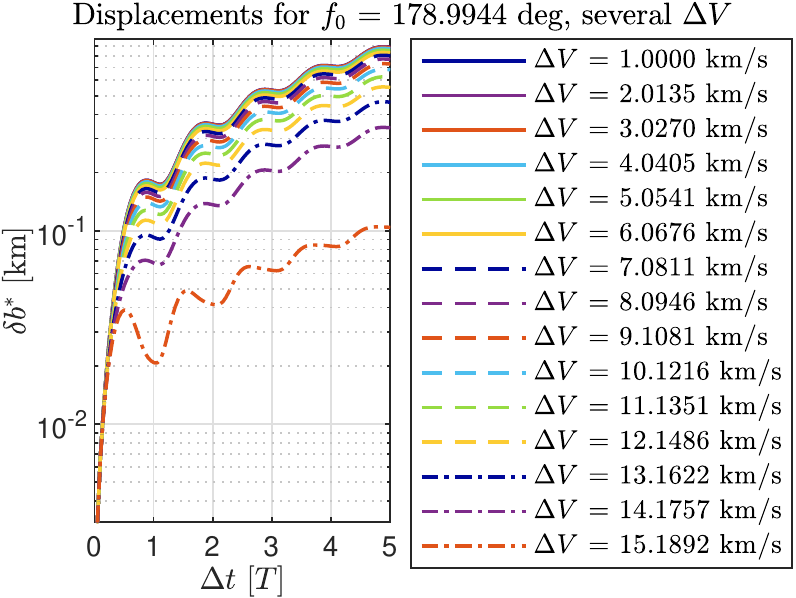}\hfill
	\includegraphics[width=0.49\textwidth]{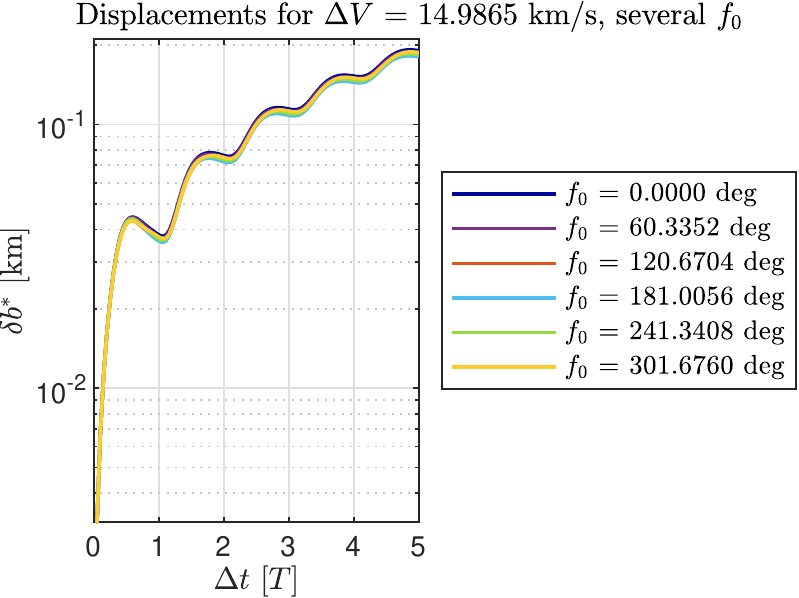}
	\caption{Maximum impact parameter for PROBA-2, for a CAM of 1~cm/s and several values of $f_0$ and $\Delta V$.}\label{fig:PROBA2_dbmax}
\end{figure}

\begin{figure}
	\centering
	\includegraphics[width=0.49\textwidth]{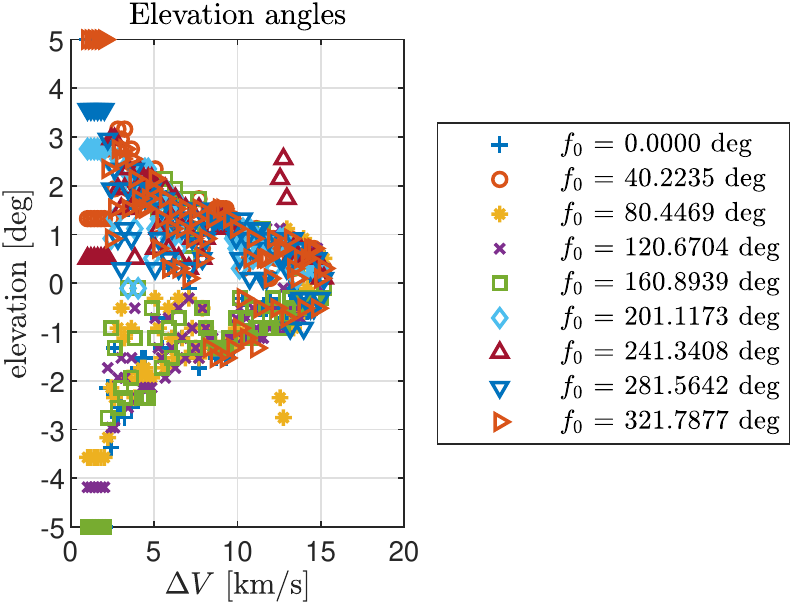}\hfill
	\includegraphics[width=0.49\textwidth]{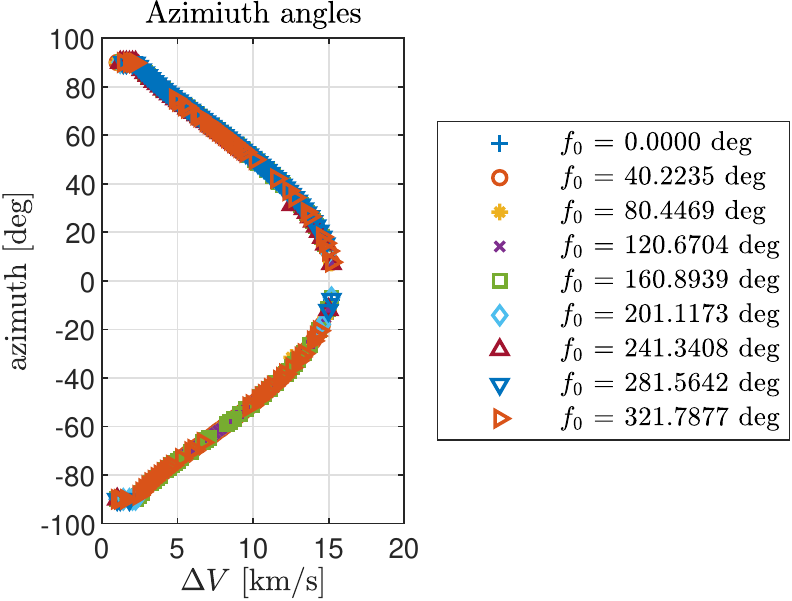}
	\caption{Elevation and azimuth angles at CA for the PROBA-2 maximum $\delta b^*$ test case.}\label{fig:PROBA2_db_angles}
\end{figure}

\begin{figure}
	\centering
	\includegraphics[width=0.60\textwidth]{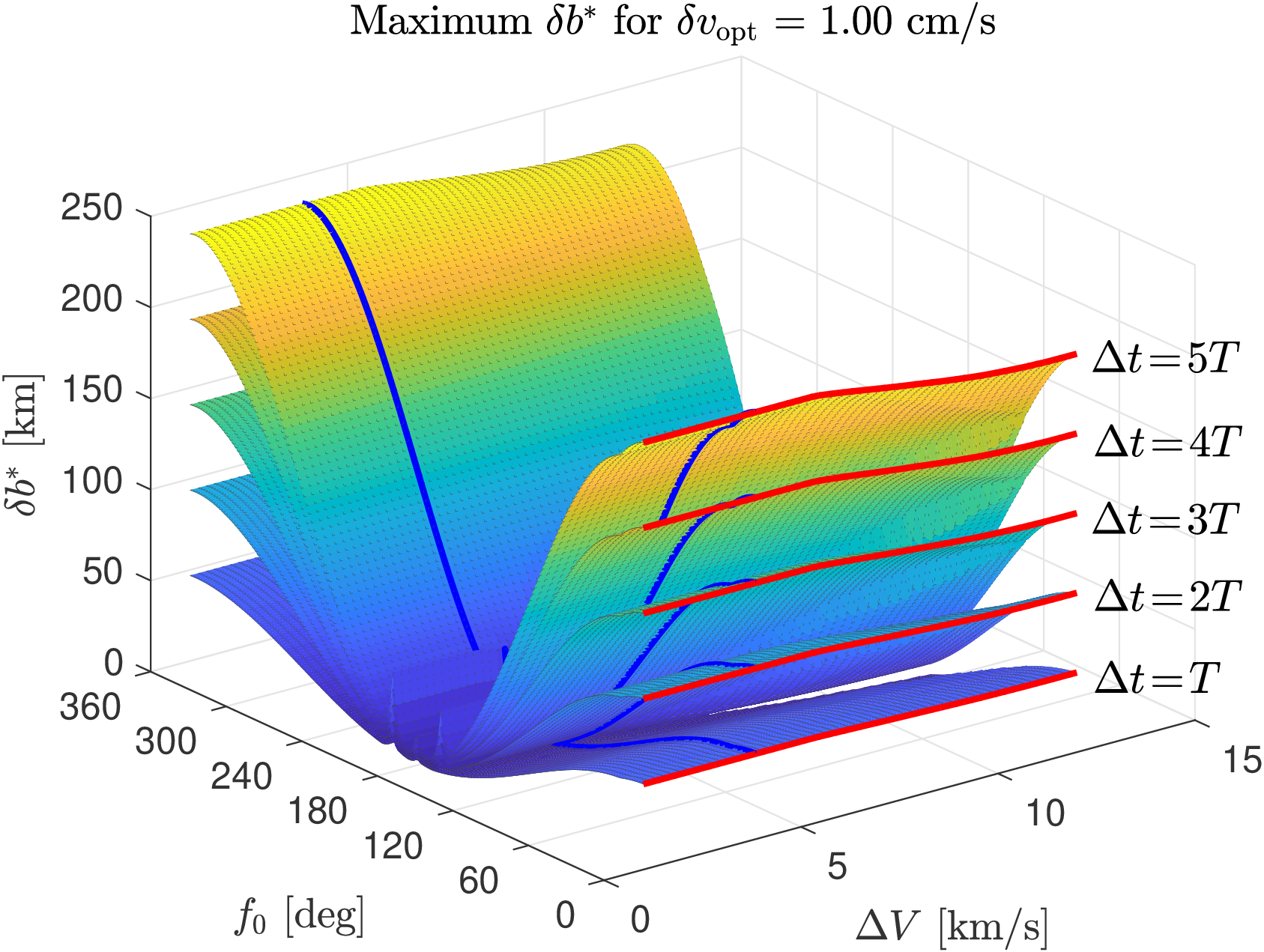}
	\caption{Maximum impact parameter for XMM for an impulsive maneuver of  1~cm/s.}\label{fig:XMM_db_surf}
\end{figure}

\begin{figure}
	\centering
	\includegraphics[width=0.49\textwidth]{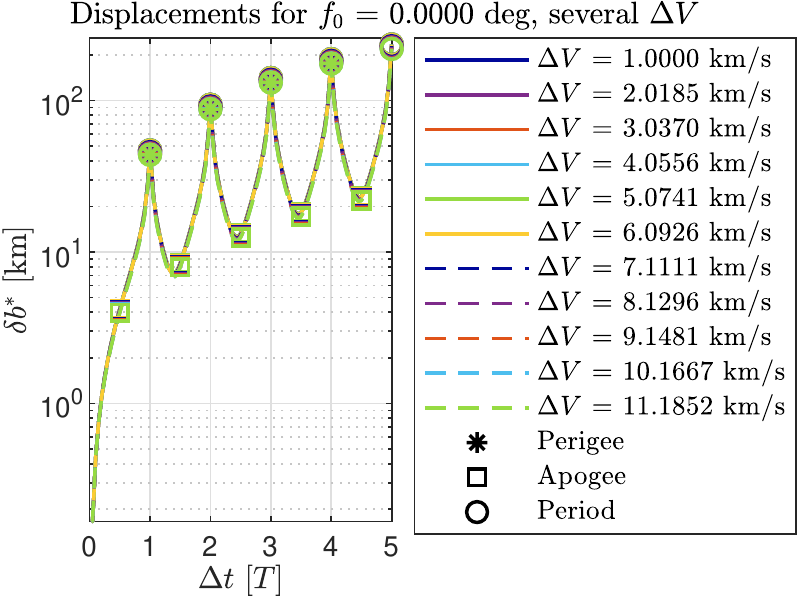}\hfill
	\includegraphics[width=0.49\textwidth]{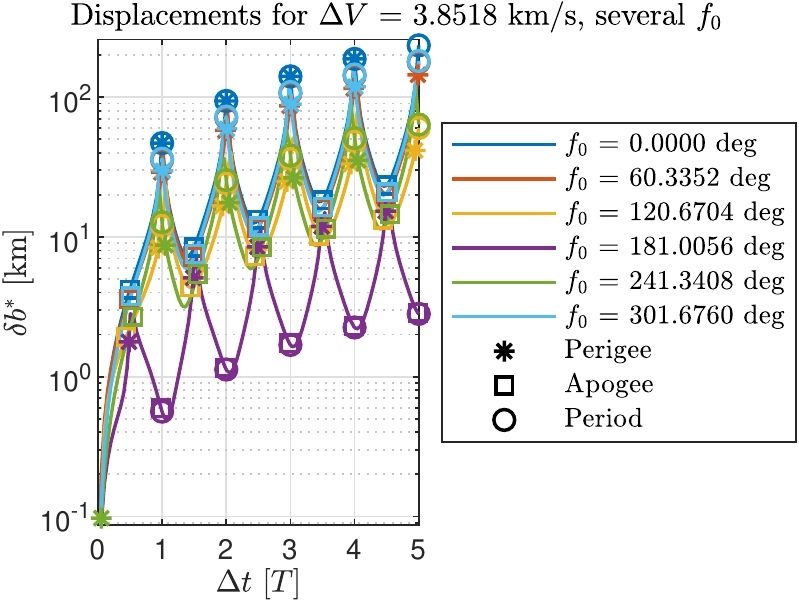}
	\caption{Maximum impact parameter for XXM, for a CAM of 1~cm/s and several values of $f_0$ and $\Delta V$.}\label{fig:XMM_dbmax}
\end{figure}

Figures~\ref{fig:PROBA2_db_surf}-\ref{fig:XMM_dbmax} present the results for the maximum impact parameter CAM, identifying some key differences with respect to the maximum $\delta r$ cases. Particularly, as it can be seen in Figs.~\ref{fig:PROBA2_db_surf} and \ref{fig:XMM_db_surf}, the deflection in the b-plane is now influenced by the geometry of the CA (represented by $\Delta V$). This effect is especially strong in the PROBA-2 (quasi-circular) test case in Figs.~\ref{fig:PROBA2_db_surf} and \ref{fig:PROBA2_dbmax}, where $\delta b^*$ goes to zero as $\Delta V$ approaches $15~\mathrm{km/s}$. The reason lies in the evolution of the azimuth and elevation angles depicted in Fig.~\ref{fig:PROBA2_db_angles}, showing that the conjunction geometry becomes closer to a head-on collision as $\Delta V$ increases. For a head-on collision trying to move along the time axis in the b-plane becomes ineffective, requiring a less efficient displacement in the geometry axis. The effect of $\Delta V$ in the elliptic case is much less significant as seen in Fig.~\ref{fig:XMM_dbmax}, but still slightly greater than for the maximum $\delta r$ results. On the other hand, the phasing of the CA and the maneuver with respect to the perigee of the spacecraft's orbit plays a very important role for the eccentric case, while the quasi-circular case now shows a small dependence with $f_0$.


\begin{figure}
	\centering
	\includegraphics[width=0.49\textwidth]{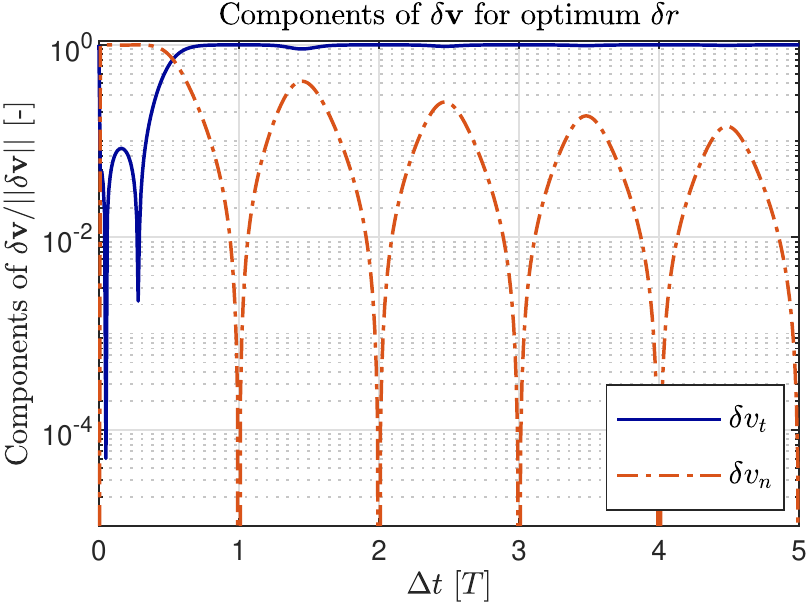}\hfill
	\includegraphics[width=0.49\textwidth]{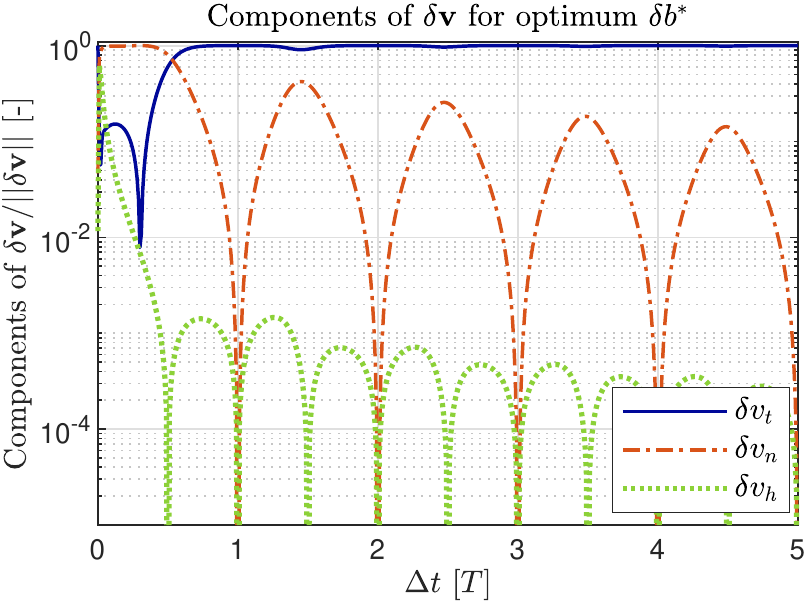}
	\caption{Optimal impulsive CAM components for the XMM test case.}\label{fig:XMM_dv}
\end{figure}

\begin{figure}
	\centering
	\includegraphics[width=0.95\textwidth]{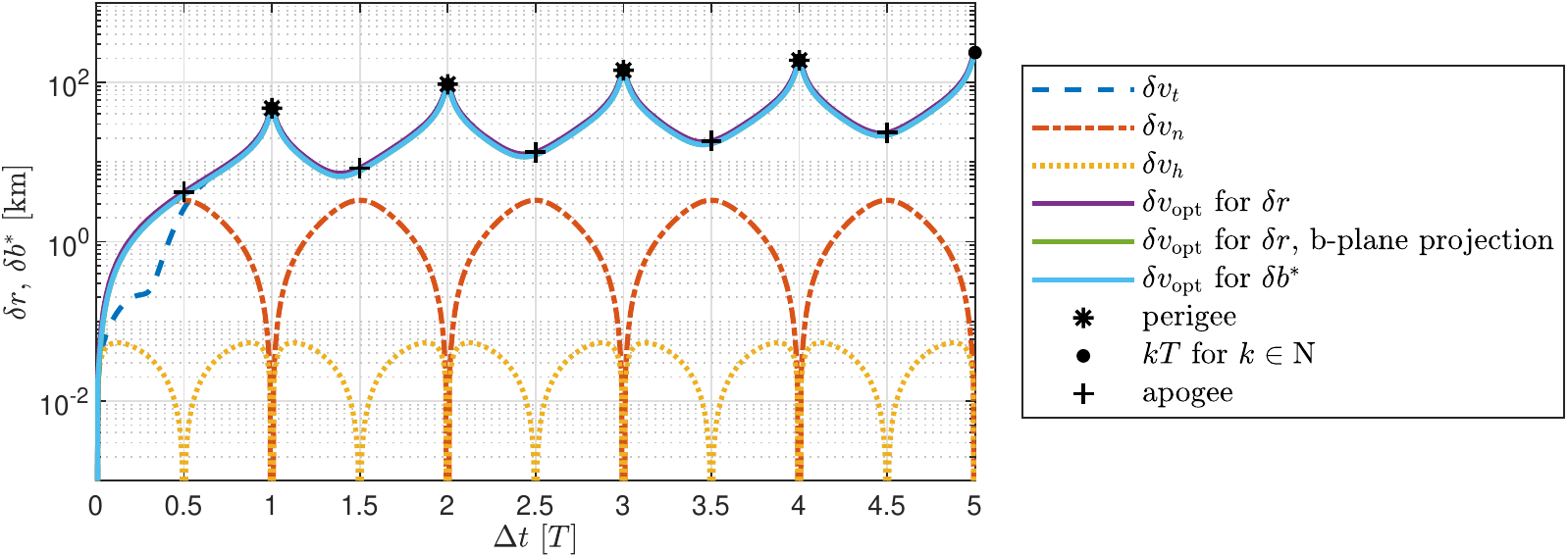}
	\caption{Displacements for several $\delta \mathbf{v}$ orientations and a $\delta v$ of 1~cm/s.}\label{fig:XMM_disp}
\end{figure}

The components in the TNH frame of the $\delta \mathbf{v}$ for maximum $\delta r$ and maximum $\delta b^*$ are represented in Fig.~\ref{fig:XMM_dv}, for the XMM test case. The true anomaly of the spacecraft at the CA is $f_\mathrm{CA} = 0~\mathrm{deg}$, whereas the debris encounters it with a relative velocity of $\Delta V = 5.0741~\mathrm{km/s}$, an elevation of $-37.5000~\mathrm{deg}$, and an azimuth of $-62.5000~\mathrm{deg}$. Same as before, in the linearized model the magnitude of the optimal $\delta \mathbf{v}$ does not affect its orientation. In both CAMs, the normal direction is dominant during the first period, while the tangential direction overcomes it for longer times. The out-of-plane component is generally negligible, and it has been omitted in the plot for the maximum $\delta r$ CAM for clarity (it falls below the range currently displayed). It is appreciably larger for the maximum $\delta b^*$ CAM, but still two orders of magnitude below the normal component. The displacements for different $\delta \mathbf{v}$ orientations and $\delta v = 1~\mathrm{cm/s}$ are shown in Fig.~\ref{fig:XMM_disp}, including both optimal CAMs as well as the results of thrusting only along the tangential, normal, and out-of-plane directions. It is observed that the maximum $\delta r$, maximum $\delta b^*$, and tangential $\delta \mathbf{v}$ solutions become very similar for lead times greater than half a period of the maneuvering spacecraft. Note that, depending on the test case, a separation can appear between the curves for $\delta r$ and $\delta b^*$, associated to the component of $\delta r$ along the direction perpendicular to the b-plane. On the other hand, the normal and out-of-plane $\delta \mathbf{v}$ orientations lead to a bounded and periodic behavior for the displacement, underlying their incapacity to leverage increases in the lead time; this justifies the fact that the optimal CAM tends to align with the tangential direction for increasing lead times.

\begin{figure}
	\centering
	\includegraphics[width=0.49\textwidth]{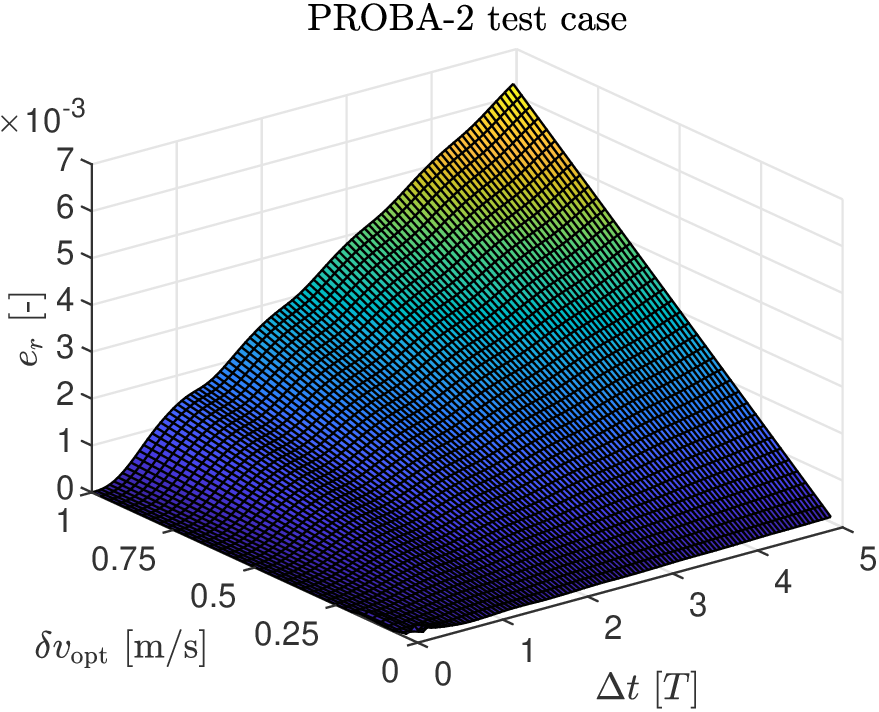} \hfill
	\includegraphics[width=0.49\textwidth]{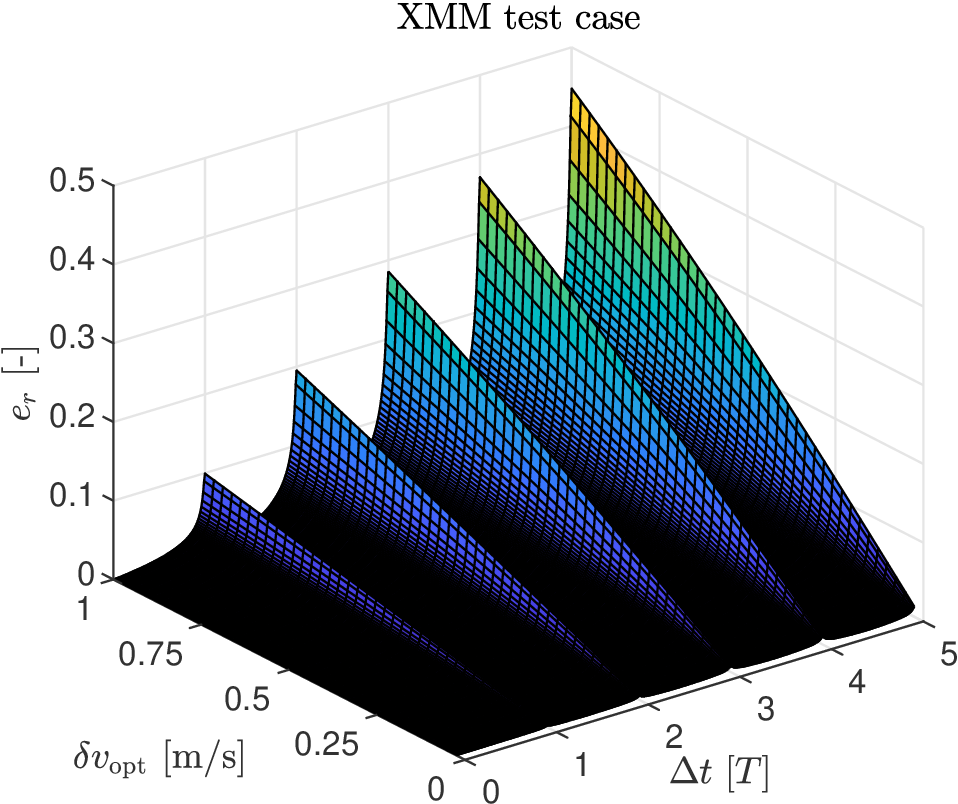}
	\caption{Relative errors in the deviation for the maximum miss distance CAM, for PROBA-2 (left) and XMM (right).}\label{fig:PROBA2_XMM_err_dr}
\end{figure}

To conclude this part of the study, the accuracy of the linearized relative motion approximation is assessed by comparing it with a high-precision numerical propagation. A relative error in the displacement is defined as~\cite{vasile2008optimal}:
\begin{equation}
e_r = \frac{|| \delta \mathbf{r}_\text{propagated} - \delta \mathbf{r}_\text{analytical}||}{||\delta \mathbf{r}_\text{propagated}||} \, ,
\end{equation}
where $\delta \mathbf{r}_\text{propagated}$ is the deviation of the spacecraft computed by the numerical propagation, and $\delta \mathbf{r}_\text{analytical}$ is the deviation given by the analytical solution. Figure~\ref{fig:PROBA2_XMM_err_dr} shows the relative errors as a function of $\delta v_\text{opt}$ and $\Delta t$ for the maximum $\delta r$ CAM, both for the PROBA-2 and XMM test cases. The nominal CA for XMM is the same already considered in Figs.~\ref{fig:XMM_dv} and \ref{fig:XMM_disp}, whereas for PROBA-2 the true anomaly of the spacecraft at the CA is $f_\mathrm{CA} = 0~\mathrm{deg}$ and the debris encounters it with a relative velocity of $\Delta V = 5.0541~\mathrm{km/s}$, an elevation of $-2.1429~\mathrm{deg}$, and an azimuth of $-73.9779~\mathrm{deg}$. In all cases, the optimal impulse orientation for maximum miss distance is determined using the linearized formulation, then $\delta \mathbf{r}_\text{propagated}$ and $\delta \mathbf{r}_\text{analytical}$ are evaluated for this direction and the desired $\delta v_\mathrm{opt}$, and finally the corresponding $e_r$ is computed. Because we are evaluating the error in the total deviation $\delta r$ for the maximum deviation CAM, the results are not actually dependent on the nominal CAs, which are reported here for completeness. The situation would be different for the impact parameter and the maximum impact parameter CAM, which depend on the conjunction geometry. It is observed that the relative error grows with the maneuver lead time and with $\delta v_\text{opt}$, because the accuracy of the relative motion approximation decreases as the separation between the nominal and modified orbits increases. The evolution with lead time shows a periodic behavior, with lines of local maxima located around the pericenters. This effect is strongly influenced by eccentricity, being barely appreciable for the quasi-circular test case but leading to steep local maxima for the highly eccentric one. This is related to the higher deviations achievable with the same impulse magnitude as eccentricity increases, and was already observed for the application of linearized relative motion to asteroid deflection by Vasile et al.~\cite{vasile2008optimal}. Compared to their results, the CAM case shows higher relative errors for similar values of $\delta v_\mathrm{opt}$; although the lead times are orders of magnitude smaller than in the asteroid case, the gravitational pull of the primary is also smaller, leading to larger displacements compared to the nominal orbit which reduce the accuracy of the linearized formulation. Finally, although the relative error for the XMM test case reaches large values between 0.4 and 0.5 for lead times of 5 periods and impulses close to $1~\mathrm{m/s}$, this does not limit the practical applicability of the method. On the one hand, most active satellites follow orbits with small eccentricities, for which $e_r$ is small. On the other hand, because the error magnitude is related to the displacement with respect to the nominal orbit, high error regions in Fig.~\ref{fig:PROBA2_XMM_err_dr} are associated to conjunction geometries for which large displacements are achievable with small values of $\delta v_\mathrm{opt}$, recall Figs.~\ref{fig:XMM_dr_surf} and~\ref{fig:XMM_db_surf}, effectively bounding the error for maneuvers with fixed displacement rather than fixed impulse magnitude.

\subsection{Minimum collision probability CAMs. Effect of uncertainties}
The analysis in the previous section does not take into account the effect of uncertainties. Indeed, although the miss distance (or impact parameter) for a fixed $\delta v$ can be easily increased by considering a longer lead time, this will also increase the uncertainties about the objects' position and velocity at the CA. This raises the question of what is the net effect of CAM lead time on collision probability. Furthermore, the maximum deviation CAM will not necessarily correspond to the minimum collision probability one for a given $\delta v_\text{opt}$. For CA involving objects with large envelopes and uncertainties, such as sails or rocket bodies, designing the CAM to minimize collision probability can prove more practical from an operational point of view. To address these key issues, maximum impact parameter and minimum collision probability CAMs are studied and compared.

The publicly available information about actual covariance matrices and CAs is very limited, as they are normally provided only to satellite operators through private CDM/CSM. For this test case, a sample covariance matrix numerically constructed from TLEs for an Iridium 33 debris (NORAD ID 33874) is used. The sample covariance has been generated through an orbit determination process based on the use of the SGP4 analytical propagator. More specifically, a uniform time grid was adopted, covering an interval of 24 hours centered at the epoch of the TLE of the object with a time step of 15 minutes. SGP4 was used to generate the state vectors of the object in Earth-centered inertial (ECI) reference frame at the epochs of the time grid. Then, a batch least-squares estimator was run to fit the generated vectors using the high-fidelity propagator AIDA~\cite{morselli2014high}. AIDA is based on the numerical integration of the dynamics of Earth-orbiting objects, including:
\begin{itemize}
\item the gravitational model EGM2008, up to the order specified by the user (order 10 was adopted in this work),
\item the atmospheric drag with the atmosphere model NRLMSISE-00 to compute air density,
\item third body perturbations (Moon and Sun),
\item solar radiation pressure with a dual-cone model for Earth shadow.
\end{itemize}
The output of the estimation process is the mean state vector and the covariance of the state of the object at the epoch of the TLE, expressed in the ECI reference frame:
\begin{gather*}
\mathbf{r}_\text{ref}^\text{ECI} = \left[ \begin{array}{ccc} +6.9688 \,\mathrm{E}{+3} & +2.0931 \,\mathrm{E}{+3} & -8.0909 \,\mathrm{E}{+0} \end{array} \right] \, \mathrm{km} \\
\mathbf{v}_\text{ref}^\text{ECI} = \left[ \begin{array}{ccc} -1.5353\,\mathrm{E}{-1} & +4.4754\,\mathrm{E}{-1} & +7.3566 \,\mathrm{E}{+0} \end{array} \right] \, \mathrm{km/s}
\end{gather*}
\begin{equation*}
\left.\mathbf{C}\right|_\text{ref} = \left[ \begin{array}{cccccc}
+1.1555 \,\mathrm{E}{-2} & -2.3144 \,\mathrm{E}{-3} & -1.1732 \,\mathrm{E}{-3} & +4.5253 \,\mathrm{E}{-7} & -5.6796 \,\mathrm{E}{-7} & -1.0945 \,\mathrm{E}{-5} \\
-2.3144 \,\mathrm{E}{-3} & +1.9147 \,\mathrm{E}{-2} & +1.4167 \,\mathrm{E}{-2} & -1.2286 \,\mathrm{E}{-5} & -2.5535 \,\mathrm{E}{-6} & -3.3049 \,\mathrm{E}{-6} \\
-1.1732 \,\mathrm{E}{-3} & +1.4167 \,\mathrm{E}{-2} & +3.0870 \,\mathrm{E}{-1} & -2.8750 \,\mathrm{E}{-4} & -8.6188 \,\mathrm{E}{-5} & -1.2493 \,\mathrm{E}{-6} \\
+4.5253 \,\mathrm{E}{-7} & -1.2286 \,\mathrm{E}{-5} & -2.8750 \,\mathrm{E}{-4} & +2.8851 \,\mathrm{E}{-7} & +7.9940 \,\mathrm{E}{-8} & +1.1511 \,\mathrm{E}{-9} \\
-5.6796 \,\mathrm{E}{-7} & -2.5535 \,\mathrm{E}{-6} & -8.6188 \,\mathrm{E}{-5} & +7.9940 \,\mathrm{E}{-8} & +4.5997 \,\mathrm{E}{-8} & +1.4570 \,\mathrm{E}{-9} \\
-1.0945 \,\mathrm{E}{-5} & -3.3049 \,\mathrm{E}{-6} & -1.2493 \,\mathrm{E}{-6} & +1.1511 \,\mathrm{E}{-9} & +1.4570 \,\mathrm{E}{-9} & +1.2022 \,\mathrm{E}{-8}
\end{array} \right] \, ,
\end{equation*} 
with units of km and km/s for length and velocity, respectively.

The sample covariance matrix is given at a particular orbital position, that is, it is associated to a specific true anomaly $f_\mathrm{ref}$. However, in order to perform a sensitivity analysis on the effect of CAM lead time the covariance matrix at the maneuver time is needed. To address this limitation, a procedure is devised to translate the reference covariance matrix to an arbitrary true anomaly. Although the reference covariance could be propagated from $f_\mathrm{ref}$ to the new true anomaly directly, this would not only update the orientation of the covariance ellipsoid but also affect its size. The aim is to preserve the size of the sample covariance ellipsoid, represented by its eigenvalues, while updating only its orientation, represented by its eigenvectors. To this end, the following procedure is proposed:
\begin{itemize}
\item The eigenvectors (i.e. principal directions) $\mathbf{e}_k$ and eigenvalues (i.e. principal values) $\lambda_k$ of $\left.\mathbf{C}\right|_\mathrm{ref}$ are computed.
\item $\left.\mathbf{C}\right|_\mathrm{ref}$ is propagated from $f_\mathrm{ref}$ to the desired true anomaly $f^*$, and its new eigenvectors $\mathbf{e}_k^*$ and eigenvalues $\lambda_k^*$ are computed.
\item The new covariance matrix at $f^*$ is retrieved by applying the eigenvalues at $f_\mathrm{ref}$, $\lambda_k$, to the eigenvectors at $f^*$, $\mathbf{e}_k^*$.
\end{itemize}
Keep in mind that the sample covariance has been constructed for testing purposes from publicly-available TLE data. The covariance matrices in an actual CDM/CSM can be appreciably smaller, but this does not affect the performance of the method or the qualitative analysis.

A sensitivity analysis is now performed for a CA taken from the previous PROBA-2 test case. The nominal Keplerian elements of spacecraft and debris at CA are reported in Table~\ref{tab:uncertainties_test_case}, corresponding to a direct impact (zero distance at CA). Both objects are assigned a covariance matrix, constructed by taking $\left.\mathbf{C}\right|_\text{ref}$ as base and applying the procedure outlined in the previous paragraph to adjust for the change in true anomaly.
\begin{table}
\centering
\caption{Spacecraft and debris nominal Keplerian elements}\label{tab:uncertainties_test_case}
\begin{tabular}{ccccccc}
\hline\hline
Object & $a$ [km] & $e$ [-] & $i$ [deg] & $\Omega$ [deg] & $\omega$ [deg] & $f_0$ [deg] \\
\hline
PROBA-2 & 7093.637 & 0.0014624 & 98.2443 & 303.5949 & 109.4990 & 179.4986 \\
Debris & 7782.193 & 0.0871621 & 88.6896 & 142.7269 & 248.1679 & 1.2233 \\
\hline\hline
\end{tabular}
\end{table}

\begin{figure}
	\centering
	\includegraphics[width=0.49\textwidth]{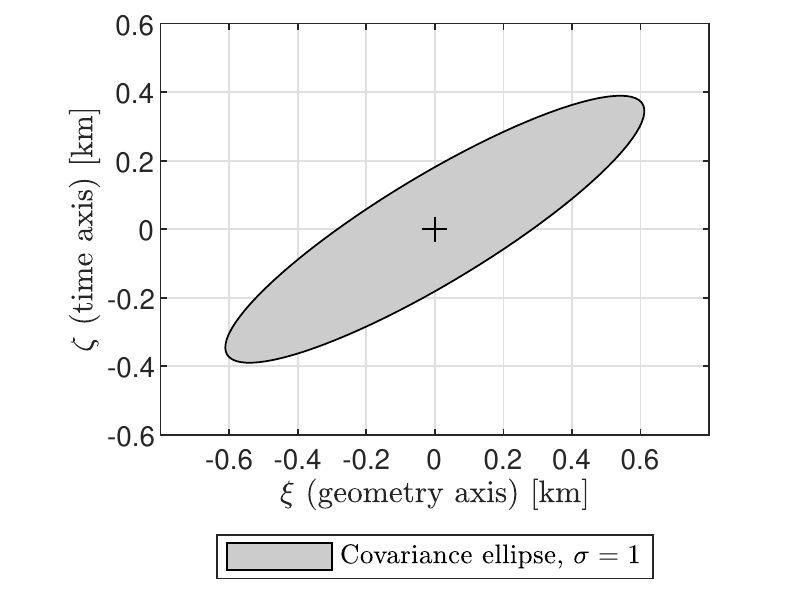} \hfill
	\includegraphics[width=0.49\textwidth]{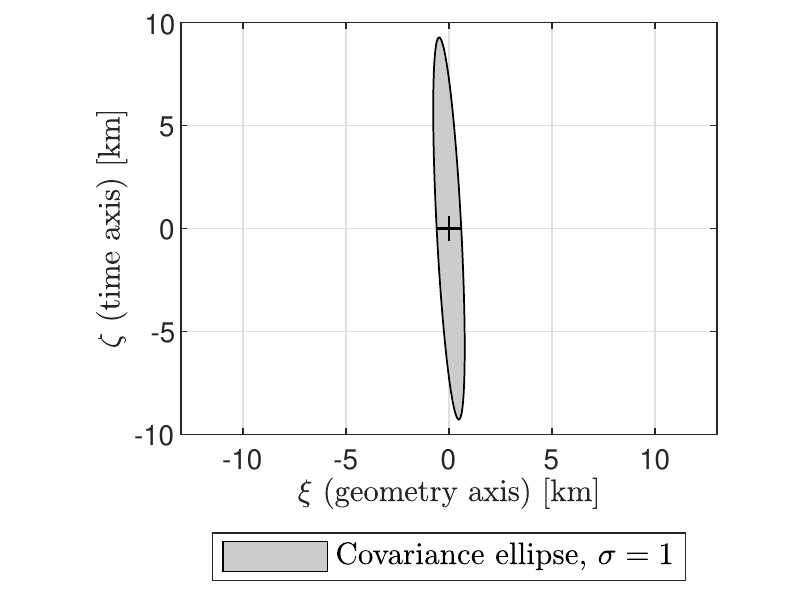}
	\caption{Combined covariance ellipse in the b-plane at CA, with orbit determination at CA (left) and 5 periods before (right).}\label{fig:cov_bpl}
\end{figure}

\begin{figure}
	\centering
	\includegraphics[width=0.49\textwidth]{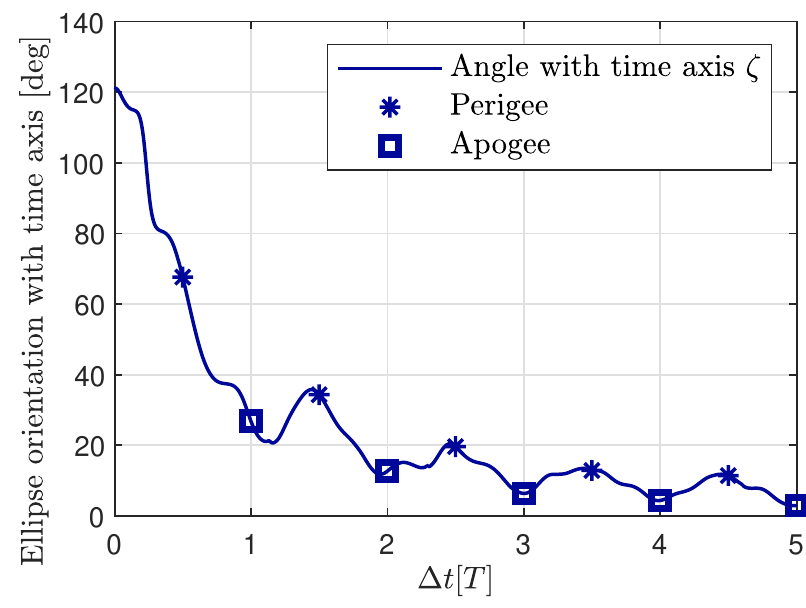}
	\caption{Orientation of the combined covariance ellipse in the b-plane with respect to the time axis.}\label{fig:ellipsoid_bpl_angle}
\end{figure}

The effect of CAM lead time in the uncertainties is evaluated first, by assuming that orbit determination for each object is performed at maneuver time, and propagating the corresponding covariance matrix up to the CA using the analytical STM introduced in Eq.~\eqref{eq:STM_full}. This is not the usual scenario for satellite operators, who are normally provided the predicted uncertainties at the CA through CDMs/CSMs. However, this artificial set-up will allow us to gain a better insight on the limitations to collision risk reduction with long lead times due to the opposing effects of increasing achievable displacements and growing uncertainties. Figure~\ref{fig:cov_bpl} shows the combined covariance ellipse in the b-plane, both for orbit determination at the CA (no lead time) and for a lead time of 5 orbital periods of the maneuvering spacecraft. As expected, introducing a lead time increases the size of the ellipse, but this does not occur in an isotropic manner. Effectively, the ellipsoid tends to grow along the time axis, causing its semi-major axis to align with it. As previously indicated, in the b-plane representation phasing-related displacements translate into displacements along the time axis, simplifying their visualization and interpretation. Then, as the time between orbit determination and covariance evaluation increases, phasing-related effects on uncertainty growth become dominant and align the principal axis of the covariance with the time axis. This is confirmed by Fig.~\ref{fig:ellipsoid_bpl_angle}, where the evolution with $\Delta t$ of the angle between the ellipse's principal direction and the time axis is represented. Angles are measured counter-clockwise in the $\xi-\zeta$ plane (that is, corresponding to rotations around the negative $\eta$ direction). The angles approach $0$ as the lead time increases, with periodic oscillations leading to local maxima around the perigee and local minima around the apogee after the first period.

\begin{figure}
	\centering
	\includegraphics[width=0.49\textwidth]{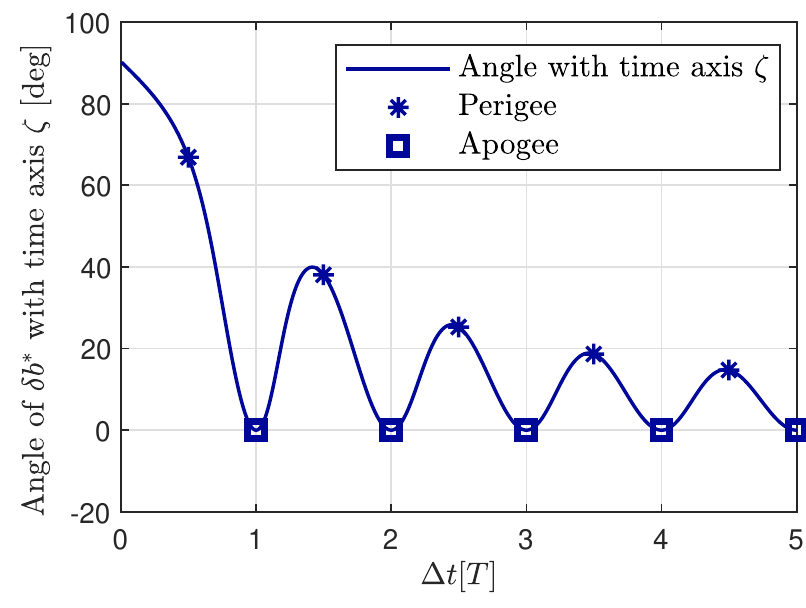}
	\caption{Orientation of the displaced impact parameter in the b-plane with respect to the time axis.}\label{fig:maxdb_bpl_angle}
\end{figure}

The tendency of the covariance ellipse to grow along and align with the $\zeta$ axis is justified by the representation of dynamics in the b-plane, where the time axis $\zeta$ corresponds to the change in phasing and the geometry axis $\xi$ to the orbit shape modification. Consequently, it is expected that the maximum impact parameter CAM will also tend to align with the $\zeta$ axis as lead time increases. This behavior is verified in Fig.~\ref{fig:maxdb_bpl_angle}, showing the angle formed in the b-plane by the deviation vector and the time axis.

\begin{figure}
	\centering
	\includegraphics[width=\textwidth]{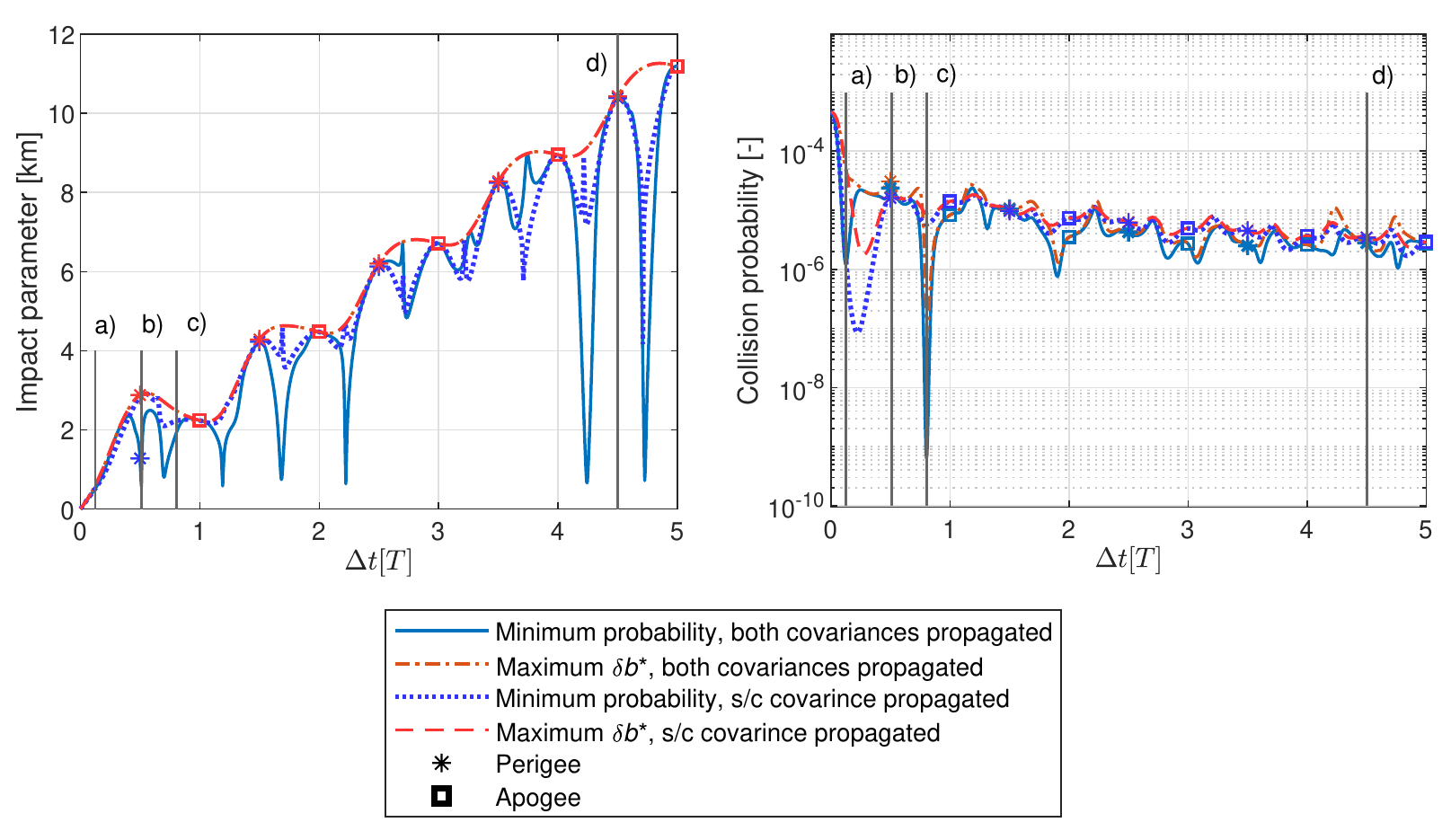}
	\caption{Impact parameter (left) and collision probability (right), for maximum impact parameter and minimum collision probability CAMs.}\label{fig:dist_prob_bothcovsVSsccov}
\end{figure}

\begin{figure}
	\centering
	\includegraphics[width=0.49\textwidth]{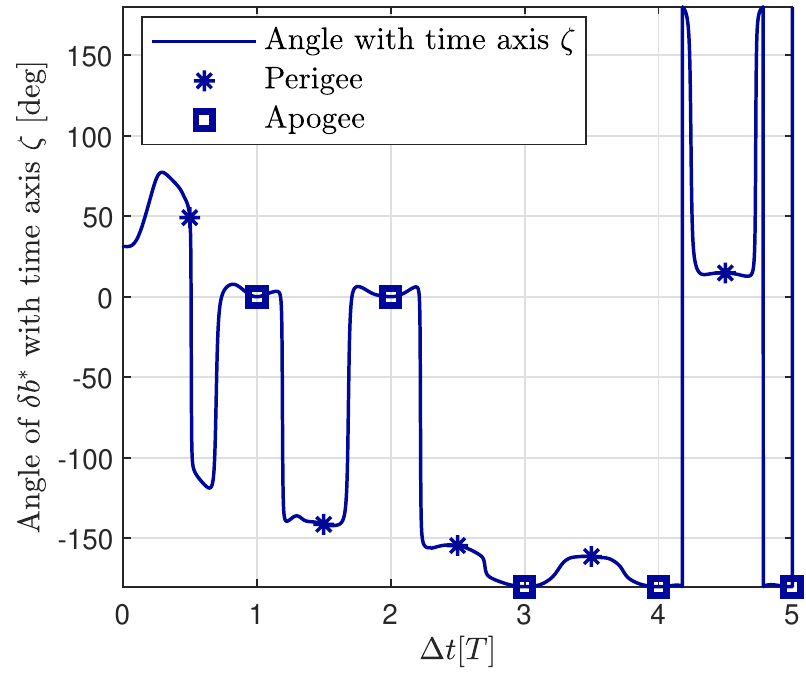}
	\caption{Orientation of the minimum collision probability CAM in the b-plane with respect to the time axis.}\label{fig:minprob_bpl_angle}
\end{figure}

The fact that both uncertainty and miss distance grow along the same direction in the b-plane casts doubts about the efficiency of the maximum deviation approach to reduce collision probability. A better solution can be sought for by trying to minimize collision probability directly. The minimum collision probability CAM is expected not only to try to increase the miss distance, but also to orient the deviation closer to the semi-minor axis of the covariance ellipse in the b-plane. Figure~\ref{fig:dist_prob_bothcovsVSsccov} compares the results obtained for the maximum deviation and minimum collision probability CAMs, both in terms of the impact parameter and the collision probability. All cases have an impulsive $\delta v_\text{opt}$ of $0.7$~m/s, and a combined envelope radius of $r_A = 10~\mathrm{m}$. Regarding the uncertainties, two different scenarios are considered: 1) the covariance matrices of both spacecraft and debris are known at the maneuver time, and 2) the covariance of the spacecraft is known at the maneuver time but the covariance of the debris is available at the CA. In the first scenario the covariances at CA of both objects change with $\Delta t$, whereas for the second scenario only the covariance of the spacecraft does. As expected, the miss distance for the maximum impact parameter CAM is always greater than or equal to the miss distance for the minimum collision probability CAM, and conversely, the collision probability for the minimum collision probability CAM is always smaller than or equal to the collision probability for the maximum impact parameter CAM. Furthermore, the impact parameter for the maximum $\delta b^*$ CAM is the same in both scenarios, as it does not depend on the uncertainties. Focusing first on the scenario where both covariance matrices are known at maneuver time, solid and dash-dot lines in Fig.~\ref{fig:dist_prob_bothcovsVSsccov}, the differences between both CAMs vary strongly with the lead time due to the evolution of the uncertainties, with some very notable features taking place for lead times within the first period. The strong, narrow minima in collision probability during the first period are due to the fast and irregular initial evolution of the combined covariance. For some values of $\Delta t$, the combined covariance ellipse has an orientation that allows for particularly efficient minimum collision probability CAMs. The maximum deviation CAM also shows low collision probability in these regions, indicating a good alignment in the b-plane between the maximum displacement CAM and the covariance semi-minor axis. These minima do not appear for lead times greater than one period, as both the preferential direction for displacement and the principal axis of the covariance tend to align with the time axis. The presence of peaks in the first period is strongly case-dependent, as observed from the differences between the solutions propagating both covariances or only that of the spacecraft. Figure~\ref{fig:minprob_bpl_angle} confirms that the minimum collision probability solution tends to separate from the $\zeta$ axis more than the maximum impact parameter one. Although it does go to $0$ or $180~\mathrm{deg}$ at the pericenters, same as the maximum $\delta b^*$ case, the evolution between these points shows large ranges of variation, as the minimum collision probability CAM steers away from the principal direction of the combined covariance. The evolution of impact parameter and collision probability in Fig.~\ref{fig:dist_prob_bothcovsVSsccov} for the minimum collision probability CAM shows a relatively irregular behavior, due to the propagation of the uncertainties and their combination at the CA. On the other hand, the results corresponding to having the orbit determination of the spacecraft at the maneuver and the orbit determination of the debris at the CA, dotted and dashed lines in Fig.~\ref{fig:dist_prob_bothcovsVSsccov}, show a smoother variation. The largest differences with respect to the previous scenario take place during the first period, especially for the collision probability. Most notably, the first minimum in collision probability is deeper and displaced towards higher lead times (but still under half a period), and the separation in miss distance between both CAM strategies strongly decreases. Regarding the miss distance, the solution propagating only the spacecraft covariance shows fewer local minima. This is consistent with the results by Bombardelli and Hernando-Ayuso~\cite{bombardelli2015optimal}, who obtained a smoother variation for test cases where the combined covariance matrix is fully known and fixed at the CA.

Although the best CAMs in terms of collision probability correspond to $\Delta t$ in the first period, their practical applicability is hindered by operational constraints. First, a last-minute CAM carries significant risks if it is not performed correctly. Furthermore, satellites in LEO may spend several orbits without coverage from their ground control stations, limiting the windows for implementing the CAM. Finally, other figures of merit apart from $\delta v$ may be considered, such as minimizing the time the satellite spends out of operation due to the CAM or the cost to restore its nominal orbit afterward. Interestingly, Fig.~\ref{fig:dist_prob_bothcovsVSsccov} shows that collision probability remains stable and slowly decreases in average for $\Delta t$ greater than one period.


\begin{figure}
	\centering
	\includegraphics[width=\textwidth]{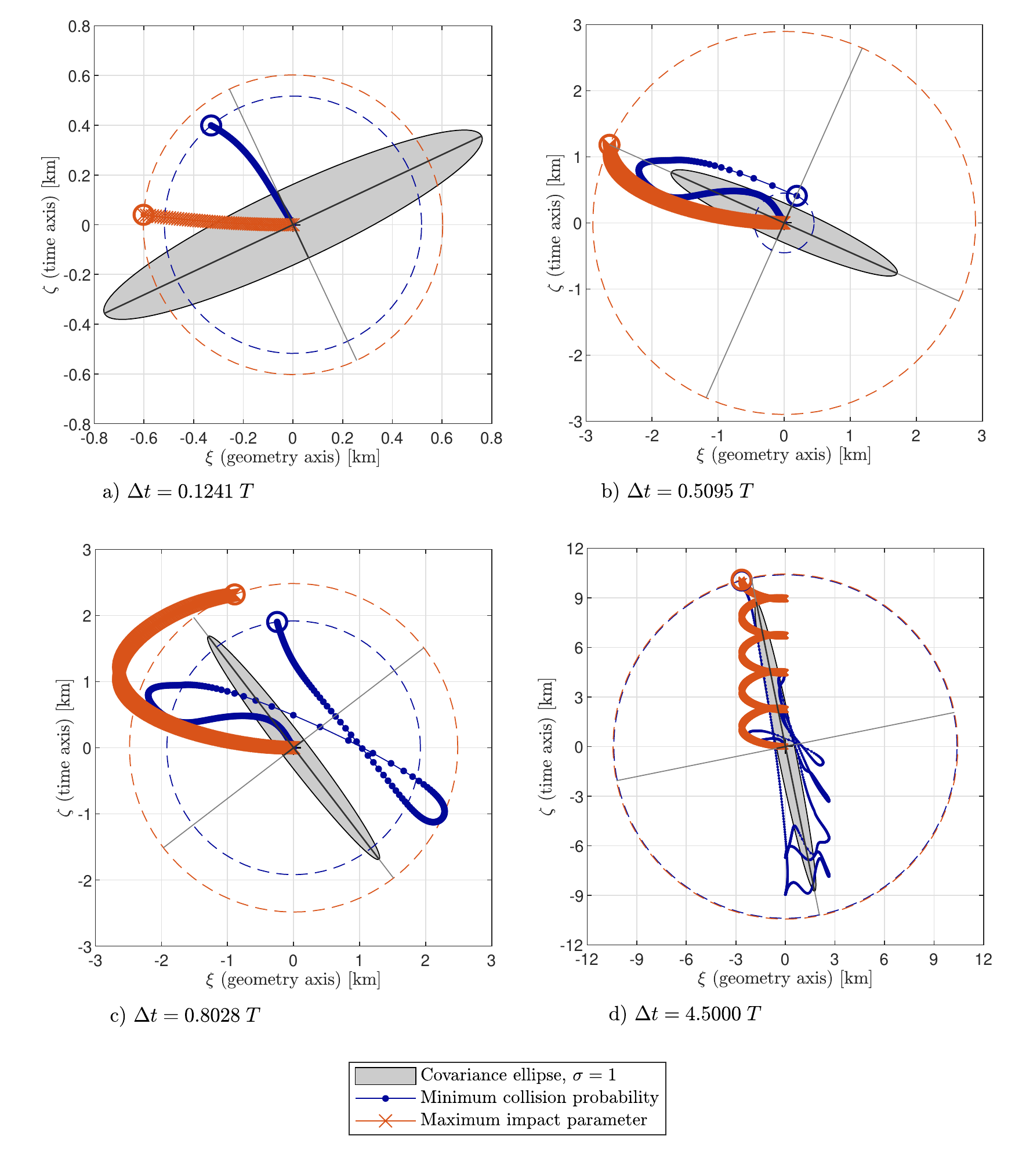}
	\caption{B-plane representation of the maximum impact parameter and minimum collision probability CAMs, for several lead times.}\label{fig:bplane_dyn}
\end{figure}

The behavior of the maximum deviation and minimum collision probability strategies can be better understood by studying the evolution of the deviated trajectories in the b-plane. Figure~\ref{fig:bplane_dyn} shows the b-plane representations of the maximum impact parameter and minimum collision probability CAMs for four different lead times, for the scenario where both covariance matrices are known at maneuver time. For convenience, these lead times are also marked in Fig.~\ref{fig:dist_prob_bothcovsVSsccov}. Three of the cases correspond to lead times smaller than one period, whereas the last one has a $\Delta t$ of 4.5 periods. As previously indicated, while CAMs within the last revolution before the encounter can provide the best performance in terms of collision probability reduction and have more dynamical interest, operational constraints for practical applications will normally require to perform the CAM several revolutions in advance. Each plot shows the 1-sigma covariance ellipsoid at CA and its principal axes for the corresponding lead time, together with the trace of the maximum $\delta b^*$ and minimum collision probability CAMs for values of $\Delta t$ between 0 and the nominal one. The points in the traces are equispaced in lead time, meaning that a higher separation between markers in the plot is associated with a faster variation of the solution as $\Delta t$ increases. To ease the visualization and comparison of the results a circle in dashed lines representing the impact parameter is included for each solution, and the final position in the b-plane is highlighted with a small circle.

The first example, depicted in Fig.~\ref{fig:bplane_dyn}a, corresponds to the first local minimum in collision probability from Fig.~\ref{fig:dist_prob_bothcovsVSsccov} ($\Delta t =0.1241\, T$). As expected, the significant difference in collision probability between both CAM strategies is due to the relative orientations of the deviated trajectories with respect to the principal directions of the covariance ellipse, with the minimum collision probability CAM closely aligned with the smallest principal direction. Interestingly, the ellipse and the maximum deviation solutions are not yet aligned with the time axis, as the lead time is short. Observing the traces, it is checked that for short lead times the maximum deviation is achieved by moving along the geometry axis, whereas the minimum collision probability CAM follows the initial direction of the semi-minor axis of the covariance ellipse (which in general will be different for each conjunction).

Figure~\ref{fig:bplane_dyn}b corresponds to the first local maximum of collision probability for the minimum collision probability CAM ($\Delta t =0.5095\, T$), which also corresponds to the first local minimum of impact parameter for this CAM, and lies in the close vicinity of the first local maximum of collision probability for the maximum deviation CAM. Same as before, the minimum collision probability solution aligns with the smallest principal direction of the covariance, although this comes at a large cost for the attainable impact parameter. In fact, its trace shows that the impact parameter is actually decreasing with $\Delta t$, and the larger separation between data points indicates that the rate of variation is increasing. Conversely, the maximum impact parameter solution lies along the principal direction of the ellipse, achieving a notably higher miss distance but with an appreciably worse collision probability. Contrary to the minimum collision probability solution, the trace for the maximum $\delta b^*$ CAM shows a smooth and uniform evolution.

The third example, see Fig.~\ref{fig:bplane_dyn}c, corresponds to the global minimum in collision probability in Fig.~\ref{fig:dist_prob_bothcovsVSsccov} ($\Delta t =0.8028\, T$). The b-plane representation shows that this global minimum is achieved through a combination of impact parameter and orientation with respect to the covariance ellipse. For longer lead times the maximum attainable deviation will keep increasing, but it will tend to closely align with the ellipsoid as previously commented. This implies that, when uncertainty evolution is taken into account, performing the maneuver sooner than the last orbital period of the spacecraft before the CA does not necessarily provide a significant advantage in terms of collision probability. Looking at the traces, the maximum $\delta b^*$ CAM follows a smooth and oscillatory behavior (recall Fig.~\ref{fig:maxdb_bpl_angle}), as it is not affected by the covariance orientation. Meanwhile, the minimum collision probability CAM shows an irregular evolution as it tries to avoid the principal axis of the moving covariance, see Fig.~\ref{fig:ellipsoid_bpl_angle}, but the final points of the trace begin to align with the time axis to leverage the higher displacements achievable through phasing.

In all the previous examples, the CAM was performed during the last revolution before the CA. While these solutions are very interesting from a dynamical perspective and provide great insight about the evolution of CAMs and uncertainties in the b-plane, their applicability in an operational scenario is limited. To address this, a case with a lead time of 4.5 periods is depicted in Fig.~\ref{fig:bplane_dyn}d. The displacements and collision probabilities are, respectively, $10.4401~\mathrm{km}$ and $2.8253\,10^{-6}$ for the maximum impact parameter CAM, and $10.3924~\mathrm{km}$ and $2.7921\,10^{-6}$ for the minimum collision probability CAM. The trace for the maximum impact parameter CAM shows a regular and periodic behavior, unaffected by the evolution of the covariance, while the minimum collision probability one displays a more irregular pattern. However, by comparing with previous solutions it is observed that the variability of the minimum $P$ CAM decreases with lead time, growing along the covariance but trying to steer clear from the principal axis. This is consistent with the results in Fig.~\ref{fig:dist_prob_bothcovsVSsccov}, where the differences between both CAMs decreased as phasing effects became dominant.

\begin{figure}
	\centering
	\includegraphics[width=0.49\textwidth]{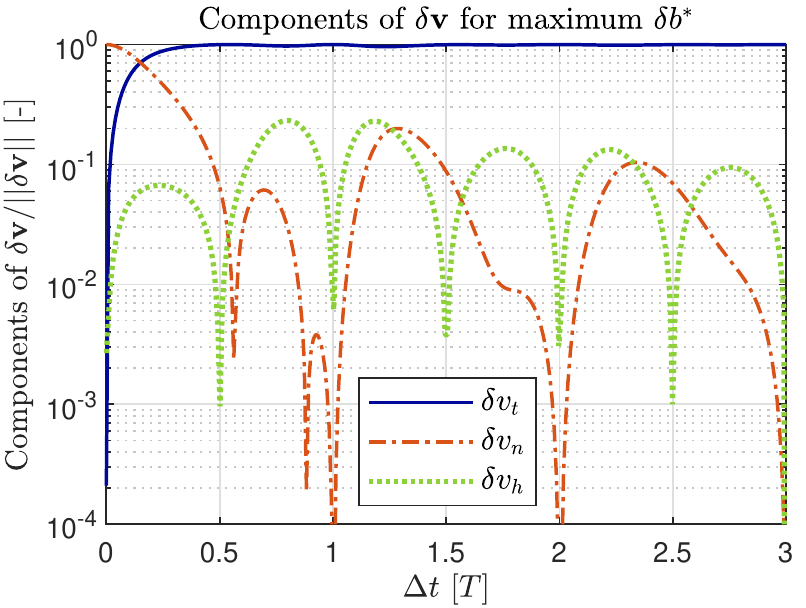} \hfill
	\includegraphics[width=0.49\textwidth]{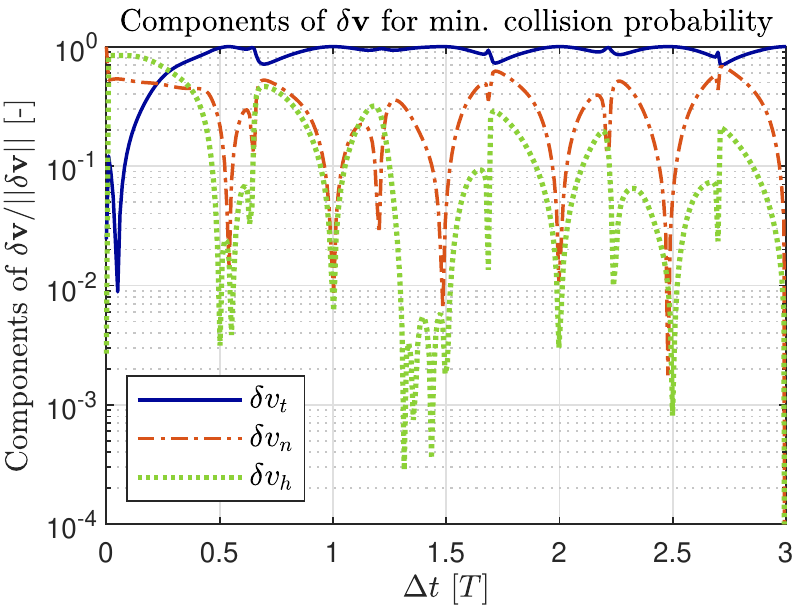}
	\caption{Optimal impulsive CAM components for maximum $\delta b^*$ (left) and minimum collision probability (right).}\label{fig:dv_tnh_uncert}
\end{figure}

The evolution of the components of $\delta \mathbf{v}$ for each type of CAM are represented in Fig.~\ref{fig:dv_tnh_uncert}, again for the scenario where both covariance matrices are known at maneuver time. Both CAMs tend to align with the tangential direction for lead times greater than a period, but the minimum collision probability one presents small deviations and higher values of the other two components. This is due to the balancing of the effects of impact parameter and in-b-plane orientation in the collision probability. Furthermore, the evolution with lead time of the uncertainties results in a less smooth short-term evolution of the control orientation.

\begin{figure}
	\centering
	\includegraphics[width=0.49\textwidth]{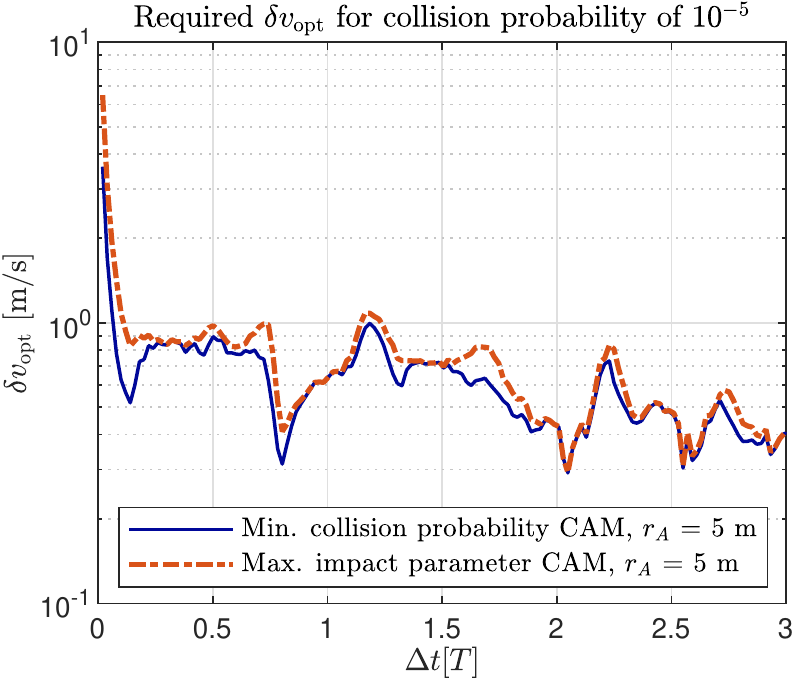} \hfill
	\includegraphics[width=0.49\textwidth]{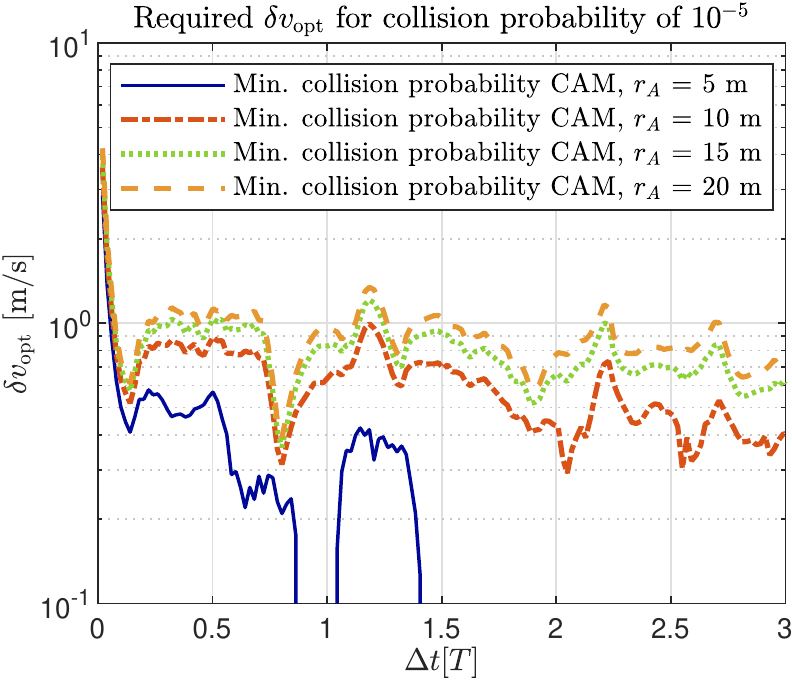}
	\caption{$\delta v_\text{opt}$ required to achieve a collision probability of $10^{-5}$, for different CAM strategies and envelope sizes.}\label{fig:req_dv_for1em5}
\end{figure}

Finally, the required $\delta v_\text{opt}$ to achieve a collision probability of $10^{-5}$ for the previous test case with both covariances known at $t_\mathrm{CAM}$ and several values of the combined envelope radius are shown in Fig.~\ref{fig:req_dv_for1em5}. As expected, the required $\delta v_\text{opt}$ is smaller for the minimum collision probability case than for the maximum $\delta b^*$ one, and it increases with the size of the envelope.

\section{Conclusions}
The design of impulsive Collision Avoidance Maneuvers (CAMs) between active spacecraft and space debris has been studied in detail, considering different control strategies and taking into account the effect of uncertainties. An analytical approximation relating the change of velocity of the spacecraft at maneuver time with the deviation at the conjunction has been applied, leveraging Gauss' planetary equations to compute the instantaneous change in orbital elements and then the linearized relative motion model to map it into a displacement at the Close Approach (CA). This model, already present in the literature, has been extended to map changes in state at a given time to changes in state at the conjunction, which allows one to propagate covariance matrices analytically. Additionally, an alternative formulation relating the impulsive CAM to changes in the impact parameter in the b-plane has been proposed.

Two different control strategies have been considered: maximum deviation and minimum collision probability. In both cases, the analytical formulation allows one to reduce the optimization problem to an eigenproblem determining the direction of the CAM. For the maximum deviation problem, two metrics have been compared: total miss distance and impact parameter in the b-plane. An extensive sensitivity analysis on the geometry of the spacecraft-debris conjunction, with ranges based on statistical data for the debris population from European Space Agency's Meteoroid and Space Debris Terrestrial Environment Reference (MASTER-2009) model, has shown that both approaches yield consistent results. However, working in the b-plane has proven more advantageous as it decomposes the deviation into effects due to the change in phasing or to the modification of the orbit geometry. Furthermore, optimizing directly for the impact parameter neglects the displacements perpendicular to the b-plane of the CA, which do not contribute to reducing the minimum distance between both objects.

As lead time for the CAM increases, both the uncertainties and the maximum deviation CAM tend to align with the time axis in the b-plane, which may limit the total decrease in collision probability from the CAM. This effect has been analyzed by comparing the maximum deviation and minimum collision probability solutions for a range of maneuver lead times. Because the aim was to analyze the opposing effects of deviation increase and uncertainty growth with lead time, the covariance matrices have been assumed to be known at maneuver time and propagated up to the CA. This is not the situation for satellite operators, who are normally provided the predicted uncertainties at the CA. The results show that minimum collision probability CAMs experience faster variations in the b-plane as they try to avoid the principal direction of the covariance, particularly during the last orbital period before the CA, which in turn leads to smaller deviations. A key observation, consistent with other works, is that the optimum $\delta \mathbf{v}$ for both CAM strategies tends to align with the tangential direction for lead times greater than half a period of the spacecraft's orbit.

The numerical test cases show that the minimum collision probability CAM outperforms the maximum deviation one in terms of propellant requirements to meet a given threshold for the collision probability. The differences are specially significant for CAMs performed during the last orbit before the CA. Although these last-minute maneuvers are normally not feasible in practical scenarios due to operational constraints, the minimum collision probability CAM still outperforms the maximum deviation one for lead times greater than a period (though the differences are reduced). It is then concluded that the minimum collision probability CAM design methodology is preferable, in concordance with the findings of previous research such as Bombardelli and Hernando-Ayuso~\cite{bombardelli2015optimal}.

The accuracy of the proposed model has been assessed for a significant range in lead times and impulse magnitudes, showing that errors remain small for practical scenarios.

\section*{Appendix: Jacobian of orbital elements with respect to position vector}

The partial derivate of $\boldsymbol\alpha = [ \; a \; e \; i \; \Omega \; \omega \; M \; ]^\top$ with respect to $\mathbf{v}$ is reported in the classic book by Battin \cite{battin1999introduction}. However, the derivation of $\partial \boldsymbol\alpha / \partial \mathbf{r}$ is not included as it is not needed to obtain Gauss' planetary equations. In this appendix we outline the derivation of $\partial \boldsymbol\alpha / \partial \mathbf{r}$ following a procedure analogous to the one in \cite{battin1999introduction}. The expressions for the partial derivatives are given in terms of orbital parameters and generic position and velocity vectors, meaning that they can be applied to any reference frame by choosing adequate expressions for $\mathbf{r}$ and $\mathbf{v}$.

A consistent ordering of vector and matrix multiplication operations is important for the developments hereafter. In the following:
\begin{equation*}
\left[ \frac{\partial \mathbf{u}}{\partial \mathbf{v} } \right]_{ij} = \frac{\partial u_i}{\partial v_j} \, ,
\end{equation*}
and vectors are treated as column vectors unless transposed.

Some equations that will be of utility are now presented. The vis-viva or energy integral takes the form:
\begin{equation}\label{eq:vis-viva}
\mu \left( \frac{2}{r} - \frac{1}{a} \right) = v^2 \, .
\end{equation}
Several relations can be established for the semilatus-rectum $p$, the semi-major axis $a$, and the semi-minor axis $b$:
\begin{equation}\label{eq:semilatus_rectum}
p = \frac{h^2}{\mu} = \frac{b^2}{a} = a \left( 1 - e^2 \right) \, .
\end{equation}
A useful relation between $\mathbf{r}$, $\mathbf{v}$, and the true anomaly $f$ is given in \cite{battin1999introduction}:
\begin{equation}\label{eq:r_dot_v}
\mathbf{r} \cdot \mathbf{v} = \frac{\mu}{h} r e \sin f \, .
\end{equation}
The partial derivative of the distance $r$ with respect to $\mathbf{r}$ takes the form:
\begin{equation}\label{eq:pderiv_r}
\pderiv{}{\mathbf{r}} r = \pderiv{}{\mathbf{r}} \sqrt{\mathbf{r}^\top \mathbf{r}} = \frac{\mathbf{r}^\top}{r} \, ,
\end{equation}
and from it the partial derivative of the position unit vector $\mathbf{i}_r$ can be obtained as:
\begin{equation}\label{eq:pderiv_ir}
\pderiv{\mathbf{i}_r}{\mathbf{r}} = \pderiv{}{\mathbf{r}} \left( \frac{\mathbf{r}}{r} \right) = \frac{1}{r} \mathbf{I}_3 - \frac{1}{r^3} \mathbf{r} \mathbf{r}^\top = - \frac{1}{r} \left( \mathbf{i}_r \mathbf{i}_r^\top - \mathbf{I}_3 \right) \, ,
\end{equation}
where $\mathbf{I}_3$ is the $3 \times 3$ identity matrix.

\subsection*{Variation of the semi-major axis}
Taking the partial derivative of the vis-viva equation, Eq.~\eqref{eq:vis-viva}, with respect to $\mathbf{r}$:
\begin{equation*}
\mu \left( -\frac{2}{r^2} \pderiv{r}{\mathbf{r}} + \frac{1}{a^2} \pderiv{a}{\mathbf{r}} \right) = \mathbf{0} \, ,
\end{equation*}
solving for the partial derivative of $a$ and introducing  Eq.~\eqref{eq:pderiv_r} leads to:
\begin{equation}\label{eq:pderiv_a_r}
\pderiv{a}{\mathbf{r}} = \frac{2 a^2}{r^3} \mathbf{r}^\top.
\end{equation}

\subsection*{Variation of the angular momentum}

Although angular momentum $h$ is not part of $\boldsymbol\alpha$, it will be needed for further derivations. The angular momentum vector $\mathbf{h}$ can be written as:
\begin{equation*}
\mathbf{h} = - \mathbf{v} \times \mathbf{r} = - \mathbf{S}_\mathbf{v} \mathbf{r} ,
\end{equation*}
where $\mathbf{S}_\mathbf{v}$ is the skew-symmetric matrix associated to operator $\mathbf{v}\times\circ$:
\begin{equation*}
\mathbf{S}_\mathbf{v} = \begin{bmatrix}
0 & -v_z & v_y \\ v_z & 0 & -v_x \\ -v_y & v_x & 0
\end{bmatrix} \, .
\end{equation*}
Using this notation, the partial derivative of $\mathbf{h}$ with respect to $\mathbf{r}$ is simply:
\begin{equation}\label{eq:pderiv_hvec_r}
\pderiv{\mathbf{h}}{\mathbf{r}} = \mathbf{S}_\mathbf{v}^\top.
\end{equation}
The partial derivative for $h$ can now be obtained by taking the partial derivative of $h^2 = \mathbf{h}^\top \mathbf{h}$ with respect to $\mathbf{r}$ and substituting Eq.~\eqref{eq:pderiv_hvec_r}, yielding:
\begin{equation}
\pderiv{h}{\mathbf{r}} = \left( \mathbf{v} \times \mathbf{i}_h \right)^\top \, ,
\end{equation}
where $\mathbf{i}_h$ is the angular momentum unit vector. This expression can be rewritten using Lagrange's formula for the expansion of the triple product and plugging in Eq.~\eqref{eq:r_dot_v}:
\begin{equation}
\pderiv{h}{\mathbf{r}} = \frac{v^2}{h} \mathbf{r}^\top - \frac{r e}{p} \sin f \mathbf{v}^\top. 
\end{equation}

\subsection*{Variation of the eccentricity}

From the definition of the semilatus rectum, Eq.~\eqref{eq:semilatus_rectum}, and recalling the previous results it is straightforward to reach:
\begin{gather*}
\pderiv{e}{\mathbf{r}} = - \frac{1}{\mu a e} \left( \mathbf{v}\times \mathbf{h}  \right)^\top + \frac{p}{ r^3 e} \mathbf{r}^\top \, ,
\end{gather*}
or alternatively, applying Lagrange's formula and Eq.~\eqref{eq:r_dot_v}:
\begin{equation}\label{eq:pderiv_e_r}
\pderiv{e}{\mathbf{r}} = \frac{1}{\mu a e} \left( \frac{h^2 a}{r^3} - v^2 \right) \mathbf{r}^\top + \frac{r \sin f}{a h} \mathbf{v}^\top \, .
\end{equation}

\subsection*{Variation of the inclination and longitude of the node}

The angular momentum vector can be expressed in the inertial frame as:
\begin{equation*}
\mathbf{h} = h \, \mathbf{i}_h = h \left( \sin\Omega \sin i \, \mathbf{i}_x - \cos\Omega  \sin i \, \mathbf{i}_y + \cos i \, \mathbf{i}_z \right) \, ,
\end{equation*}
where $\mathbf{i}_x$, $\mathbf{i}_y$, and $\mathbf{i}_z$ are the unit vectors along axes $x$, $y$, and $z$, respectively. Taking the partial derivative with respect to $\mathbf{r}$ leads to:
\begin{equation*}
\pderiv{ \mathbf{h}}{\mathbf{r}} = \mathbf{i}_h \pderiv{h}{\mathbf{r}} + h \sin i \, \mathbf{i}_n \pderiv{\Omega}{\mathbf{r}} - h \, \mathbf{i}_m \pderiv{i}{\mathbf{r}} \, ,
\end{equation*}
where $\mathbf{i}_n=\cos\Omega \mathbf{i}_x + \sin\Omega \mathbf{i}_y$ is the line of nodes unit vector, and $\mathbf{i}_m = \mathbf{i}_h \times \mathbf{i}_n$. Substituting Eq.~\eqref{eq:pderiv_hvec_r} and projecting along $\mathbf{i}_n$ leads to the partial derivative for $\Omega$:
\begin{equation}\label{eq:pderiv_RAAN_v1}
\pderiv{\Omega}{\mathbf{r}} = \frac{1}{h \sin i} \left( \mathbf{S}_\mathbf{v} \mathbf{i}_n \right)^\top = \frac{1}{h \sin i} \left( \mathbf{v} \times \mathbf{i}_n \right)^\top  \, ,
\end{equation}
whereas projecting along $\mathbf{i}_m$ yields the partial derivative for $i$:
\begin{equation}\label{eq:pderiv_i_v1}
\pderiv{i}{\mathbf{r}} = - \frac{1}{h} \left( \mathbf{S}_\mathbf{v} \mathbf{i}_m \right)^\top = - \frac{1}{h} \left( \mathbf{v} \times \mathbf{i}_m \right)^\top \, .
\end{equation}

Cross products $\mathbf{v} \times \mathbf{i}_n$ and $\mathbf{v} \times \mathbf{i}_m$ can be rewritten in a more convenient way expressing the velocity vector as \cite{battin1999introduction}:
\begin{equation*}
\mathbf{v} = - \frac{\mu}{h} \sin f \mathbf{i}_e + \frac{\mu}{h} \left( e + \cos f \right) \mathbf{i}_p \, ,
\end{equation*}
where $\mathbf{i}_e$ is the eccentricity unit vector, and $\mathbf{i}_p = \mathbf{i}_h \times \mathbf{i}_e$. Performing some manipulations one obtains:
\begin{equation*}
\mathbf{v} \times \mathbf{i}_n = - \frac{\mu}{h} \left[ \cos \theta + e \cos \omega \right] \mathbf{i}_h \, ,
\end{equation*}
\begin{equation*}
\mathbf{v} \times \mathbf{i}_m = - \frac{\mu}{h} \left[ \sin \theta + e \sin \omega \right] \mathbf{i}_h \, .
\end{equation*}
Substituting these expressions, Eqs~\eqref{eq:pderiv_RAAN_v1} and \eqref{eq:pderiv_i_v1} take the more convenient forms:
\begin{equation}\label{eq:pderiv_RAAN_v2}
\pderiv{\Omega}{\mathbf{r}} = - \frac{\cos\theta + e \cos\omega}{p \sin i} \mathbf{i}_h^\top \, ,
\end{equation}
\begin{equation}\label{eq:pderiv_i_v2}
\pderiv{i}{\mathbf{r}} = \frac{\sin\theta + e \sin\omega}{p} \mathbf{i}_h^\top \, .
\end{equation}

\subsection*{Variation of the argument of pericenter, argument of longitude, and true anomaly}

The argument of pericenter $\omega$ is related to the argument of latitude $\theta$ and the true anomaly $f$ through the expression $\theta = \omega + f$, leading to:
\begin{equation*}
\pderiv{\omega}{\mathbf{r}} = \pderiv{\theta}{\mathbf{r}} - \pderiv{f}{\mathbf{r}}.
\end{equation*}

The partial derivative for the true anomaly is obtained first. From the equation of the orbit:
\begin{equation*}
r \left( 1 + e \cos f \right) = \frac{h^2}{\mu} \, ,
\end{equation*}
it is possible to derive:
\begin{equation*}
r e \sin f \pderiv{f}{\mathbf{r}} = \pderiv{r}{\mathbf{r}} \left( 1 + e \cos f \right) + r \cos f \pderiv{e}{\mathbf{r}} - \frac{2 h}{\mu} \pderiv{h}{\mathbf{r}}.
\end{equation*}
However, solving for $\partial f / \partial \mathbf{r}$ directly from this equation would lead to singularities for $ f = 0, \pi$. On the other hand, taking the partial derivative of Eq.~\eqref{eq:r_dot_v} with respect to $\mathbf{r}$ leads to:
\begin{equation*}
r e \cos f \pderiv{f}{\mathbf{r}} = \frac{h}{\mu} \mathbf{v}^\top + \frac{\mathbf{r}\cdot\mathbf{v}}{\mu} \pderiv{h}{\mathbf{r}} - e \sin f \pderiv{r}{\mathbf{r}} - r \sin f \pderiv{e}{\mathbf{r}} \, ,
\end{equation*}
which is singular for $f=\pm\pi$. Combining both equations an expression valid for all $f$ is obtained:
\begin{equation*}
r e \pderiv{f}{\mathbf{r}} = \sin f \pderiv{r}{\mathbf{r}} + \frac{1}{\mu} \left( \cos f \mathbf{r} \cdot \mathbf{v} - 2 \sin f h \right) \pderiv{h}{\mathbf{r}} + \frac{h \cos f}{\mu} \mathbf{v}^\top \, .
\end{equation*}
Substituting for the known partial derivatives and manipulating to get a more compact form one finally reaches:
\begin{equation}
\pderiv{f}{\mathbf{r}} = \frac{r}{e h^2} \left[ \sin f \left( \frac{h^2}{r^3} - \left(r+p\right) \frac{v^2}{r^2} \right) \mathbf{r}^\top + \left( \frac{h}{p}\left( \cos f + e \right) + \frac{eh}{r} \right) \mathbf{v}^\top \right] \, .
\end{equation}
Alternatively, the norm of the velocity can be eliminated by using the vis-viva equation, Eq.~\eqref{eq:vis-viva}:
\begin{equation}\label{eq:pderiv_f_r}
\pderiv{f}{\mathbf{r}} = \frac{r}{e h^2} \left[ \frac{\mu}{a r^3} \left( (r-a)(p+r) - r a \right) \sin f \mathbf{r}^\top + \left( \frac{h}{p}\left( \cos f + e \right) + \frac{eh}{r} \right) \mathbf{v}^\top \right] \, .
\end{equation}

The partial derivative of the argument of latitude $\theta$ with respect to $\mathbf{r}$ is obtained following an anologous procedure to the one by Battin~\cite{battin1999introduction} for $\partial \theta/\partial\mathbf{v}$. We begin by expressing $\theta$ in terms of $\Omega$, $\mathbf{i}_x$, $\mathbf{i}_y$, and $\mathbf{i}_r$:
\begin{equation*}
\cos\theta = \mathbf{i}_n \cdot \mathbf{i}_r = \cos\Omega \left( \mathbf{i}_x \cdot \mathbf{i}_r \right) + \sin\Omega \left( \mathbf{i}_y \cdot \mathbf{i}_r \right) \, .
\end{equation*}
Taking the partial derivative with respect to $\mathbf{r}$, substituting for Eq.~\eqref{eq:pderiv_ir}, and operating:
\begin{equation*}
\pderiv{\theta}{\mathbf{r}} = - \cos i \pderiv{\Omega}{\mathbf{r}} + \frac{1}{r} \mathbf{i}_\vartheta^\top .
\end{equation*}
Replacing $\partial\Omega/\partial\mathbf{r}$ with Eq.~\eqref{eq:pderiv_RAAN_v2} and expressing the transversal unit vector $\mathbf{i}_\vartheta$ in terms of $\mathbf{r}$ and $\mathbf{v}$, an expression with the same structure of previous results is reached:
\begin{equation}
\pderiv{\theta}{\mathbf{r}} = \frac{(\cos\theta + e \cos\omega) \cos i}{p \sin i} \mathbf{i}_h^\top - \frac{e}{pr} \sin f \mathbf{r}^\top + \frac{1}{h} \mathbf{v}^\top .
\end{equation}

Finally, the partial derivative for the argument of pericenter $\omega$ is obtained combining the results for $f$ and $\theta$:
\begin{equation}
\pderiv{\omega}{\mathbf{r}} = \frac{(\cos\theta + e \cos\omega) \cos i}{p \sin i} \mathbf{i}_h^\top  -\frac{r}{h^2e}\sin f \left( \frac{h^2}{p r^3} (p+e^2r) - (p+r)\frac{v^2}{r^2} \right) \mathbf{r}^\top - \frac{r}{h e p} \left( \cos f + e \right) \mathbf{v}^\top \, .
\end{equation}

\subsection*{Variation of eccentric and mean anomalies for elliptic orbit}

For the particular case of elliptic orbit the eccentric anomaly $E$ is defined as:
\begin{equation*}
\cos E = \frac{ \cos f + e}{1 + e \cos f} \, .
\end{equation*}
Taking the partial derivative with respect to $\mathbf{r}$ and using relation $b \sin E = r \sin f$ (see~\cite{battin1999introduction}, section 4.3) to remove the trigonometric terms in $E$ one reaches:
\begin{equation}\label{eq:dEdr}
\pderiv{E}{\mathbf{r}} = - \sin f \frac{r a}{p b} \pderiv{e}{\mathbf{r}} + \frac{r}{b} \pderiv{f}{\mathbf{r}}.
\end{equation}
Substituting previous results, grouping in $\mathbf{r}^\top$ and $\mathbf{v}^\top$, and simplifying:
\begin{equation}
\pderiv{E}{\mathbf{r}} = \frac{r}{\mu b e} \left[ -\sin f \frac{a \mu + r\left( r v^2 - \mu \right) }{r^3} \mathbf{r}^\top + \frac{h}{p} \left( \cos f + e \right) \mathbf{v}^\top \right] \, ,
\end{equation}
or alternatively, using the vis-viva equation, Eq.~\eqref{eq:vis-viva}, to remove $v^2$:
\begin{equation}
\pderiv{E}{\mathbf{r}} = \frac{r}{\mu b e} \left[ \mu \sin f \frac{r^2-ra-a^2}{a r^3} \mathbf{r}^\top + \frac{h}{p} \left( \cos f + e \right) \mathbf{v}^\top \right] .
\end{equation}

Finally, the partial derivative for the mean anomaly can be derived from Kepler's equation:
\begin{equation*}
M = E - e \sin E .
\end{equation*}
Taking the partial derivative, plugging in Eqs.~\eqref{eq:dEdr}, \eqref{eq:pderiv_e_r} and \eqref{eq:pderiv_f_r}, and simplifying:
\begin{equation}
\pderiv{M}{\mathbf{r}} = \frac{r b}{h a^2 e} \left[ \frac{h}{p r^3} \left( r^2 - a\left(p+r\right) \right) \sin f \mathbf{r}^\top + \cos f \mathbf{v}^\top \right].
\end{equation}

\section*{Funding Sources}
The research performed for this paper has received funding from the European Research Council (ERC) under the European Union's Horizon 2020 research and innovation programme within the project COMPASS (grant agreement No 679086). The simulations have been performed within the study contract ``Environmental aspects of passive de-orbiting devices'' funded by the European Space Agency (Space Debris Office) (contract number 4000119560/17/F/MOS). 

\section*{Acknowledgments}
The authors want to thank Vitali Braun and Benjamin Bastida Virgili from European Space Agency's Space Debris Office, the three anonymous reviewers, and the Associate Editor for their useful comments and suggestions.


\end{document}